\documentclass{aa}  

\usepackage{graphicx}
\usepackage{txfonts}
\usepackage{amsmath}	
\usepackage{amssymb}
\usepackage{verbatim}
\usepackage{xcolor}

\newcommand{\xmm}{\hbox{\it{XMM-Newton}}}

\newcommand{\chandra}{\hbox{\it{Chandra}}}

\def\dn{$\rm D_{n}4000$}
\def\halpha{$\rm H{\alpha}$}
\def\hbeta{$\rm H{\beta}$}

\def\lx{$L_{\rm X}$}

\def\lo{$L_{\rm [O III]}$}

\def\mstar{$M_\star$}

\def\sfrssp{SFR$_{\rm ssp}$}

\def\sersic{$\rm S{\acute e}rsic$}
\def\sigmaone{$\Sigma_{1}$}
\def\sigmam{$\Sigma_{\star}$}
\def\sigmasfr{$\Sigma_{\rm SFR}$}
\def\sigmassfr{$\Sigma_{\rm sSFR}$}
\def\sigmasfrone{$\Sigma_{\rm SFR, 1~kpc}$}

\def\dms{$\Delta \rm MS$}

\def\re{$r_{\rm e}$}
\def\sersic{$\rm S{\acute e}rsic$}

\def\lx{$L_{\rm X}$}
\def\lxlimit{$L_{\rm X, limit}$}
\def\dn{$\rm D_{n}4000$}

\nocite{*}
\begin{document} 

   \title{Investigating central star formation in local AGN host galaxies:\\ is there tension between coeval growth and AGN feedback?}
   \subtitle{}

   \author{Q. Ni
          \inst{1}\fnmsep\thanks{qni@mpe.mpg.de},
          K. Nandra\inst{1},
          A. Merloni\inst{1},
          J. Comparat\inst{1,2},
          D. Tubín-Arenas\inst{3},
          \and
          Y. Zhang\inst{1,4}
          }
   \institute{Max-Planck-Institut f\"{u}r extraterrestrische Physik (MPE), Gie{\ss}enbachstra{\ss}e 1, D-85748 Garching bei M\"unchen, Germany
   		\and
             Univ. Grenoble Alpes, CNRS, Grenoble INP, LPSC-IN2P3, 53, Avenue des Martyrs, 38000, Grenoble, France
             \and
             Leibniz-Institut f\"{u}r  Astrophysik Potsdam (AIP), An der Sternwarte 16, 14482 Potsdam, Germany             
             \and
             Max-Planck-Institut f\"{u}r Astrophysik (MPA), Karl-Schwarzschild-Str. 1, D-85741 Garching, Germany
             }
\authorrunning{Ni et al.}
\titlerunning{The central star formation properties of local AGN host galaxies}
   \date{}

  \abstract
   {It has been argued that supermassive black holes (BHs) coevolve with the central parts of galaxies, as a result of the common fuel for both the BH and star formation in the galaxy central region, as supported by the particularly significant relation between BH growth and the central mass density within 1 kpc (\sigmaone) found among star-forming galaxies. In the context of this scenario, one would naturally expect a close observational link between AGN activity and star formation activity in the central regions, e.g., the surface star formation rate density in the central 1 kpc region (\sigmasfrone), as the manifestation of coeval growth.
With $\approx$ 3000 galaxies in the Mapping Nearby Galaxies at Apache Point Observatory (MaNGA) survey that have X-ray coverage from SRG/eROSITA, \xmm, or \chandra, we studied how the X-ray AGN fraction varies with \sigmasfrone. 
We found that the fraction of X-ray AGNs with relatively higher specific BH accretion rates increases with \sigmasfrone, consistent with the expectation.
Comparison of the mean star formation rate surface density (\sigmasfr) profiles of the host galaxies of these AGNs and normal galaxies sharing similar properties reveals elevated \sigmasfr\ in AGN hosts across the entire central region.
As for optically-selected AGNs, their hosts also tend to show high \sigmasfr\ in the central regions on average compared to normal galaxies, but are discrepant with X-ray AGNs in terms of the trend of AGN fraction vs. \sigmasfrone, which can be explained by selection effects.
While these general trends all support the coeval growth scenario, they do not contradict observational evidence for AGN feedback, as the time-averaged effects from local AGN feedback are modest in star-forming regions.
   }

   \maketitle

\section{Introduction}

Supermassive black holes (BHs) are thought to coevolve with their host galaxies, as motivated by the well-established correlation between BH mass and bulge properties in the local universe \citep[e.g.][]{KH2013}. Two processes are generally invoked in this scenario. First, because BHs and galaxies share the same gas reservoir, some level of coeval growth is naturally expected. Second, AGN feedback is considered to play an important role in the regulation of galaxy evolution \citep[e.g.][]{Croton2006} by directly interacting with the gas reservoir via different modes and channels \citep[e.g.][]{Fabian2012, Alexander2012, Harrison2017}.

It has been found that BH growth is closely related to various host-galaxy properties, and compactness in the galaxy central regions is particularly effective in tracking the long-term average BH growth level among star-forming (SF) galaxies \citep{Ni2019, Ni2021}, 
which supports the picture of gas in the central $\sim$kpc region of galaxies as a common fuel for both AGNs and galaxies. 
In this scenario, one would expect a higher AGN fraction/BH growth level among galaxies with high central star formation rate (SFR), similar to what is being observed for the global SFR \citep[e.g.][]{Chen2013, Andonie2024}.
At the same time, there have been some studies showing AGNs with suppressed gas fraction or star formation in the centers \citep[e.g.][]{Bing2019, Ellison2021, Lammers2023}, which have been utilized as evidence for AGN feedback, but are potentially in tension with the picture of coeval growth.

Integral field unit (IFU) surveys are widely utilized in studies of this kind to spatially resolve the properties of AGN hosts and compare them with those of normal galaxies (that are selected to constitute a control sample).
Currently, due to the limited number of JWST IFU observations and the limited understanding towards recovering the observations obtained to a complete and unbiased sample, this type of study has been restricted to the low-redshift universe with IFU data from CALIFA \citep{Sanchez2012}, SAMI \citep{Bryant2015}, and the MaNGA survey from Sloan Digital Sky Survey IV (SDSS-IV; \citealt{Bundy2015}), which has the largest sample size, ideal for population studies.
In different studies, AGNs are selected at different wavelengths via diverse selection criteria, and normal galaxies in the control sample are also selected differently, which might explain the wide range of (and sometimes contradictory) results arising.
When comparing optically-selected AGNs with normal galaxies, both suppressed central SFR \citep[e.g.][]{Bing2019, Lammers2023} and elevated central SFR \citep[e.g.][]{Sanchez2018, Gatto2025} are reported.
 \citet{Mulcahey2022} studied radio-detected AGNs (which are selected combining radio, optical, and mid-infrared diagnostics), and found that the average stellar population age (which can track specific SFR) profile of AGN hosts is similar to that of control galaxies.
In addition to IFU-based studies, there have also been attempts established based on multi-band pixel-by-pixel SED fitting. 
Utilizing narrowband photometric surveys, \citet{minijpas2024} report a suppression of specific SFR in the central regions of X-ray-selected AGN hosts compared with matched SF galaxies.
While AGN selection methods vary in their respective merits and drawbacks, X-ray selection is considered to be rather clean and unbiased \citep[e.g.][]{Hickox2018}.
However, due to the lack of X-ray coverage with IFU surveys, previous studies based on IFU data do not fully make use of X-ray-selected AGNs in the sample.

In this work, we combine X-ray AGN selection with optical AGN selection of MaNGA galaxies to 
further study how AGN activity links with the star formation properties in the central regions of galaxies, and particularly, trying to resolve the potential
tension of coeval growth expected from a common fuel for both the BH and the galaxy, where AGN activity is linked with high central \sigmasfr, and the so-called AGN negative feedback, where AGN activity is linked to suppressed central \sigmasfr. We also discuss how the AGN selection method might affect the observed results.

The paper is structured as follows. In Section~\ref{s-sample}, we describe the sample construction process. 
In Section~\ref{s-ar}, we detail the analysis results and discuss what they imply in Section~\ref{s-dc}. 
The conclusions are presented in Section~\ref{s-c}.
Throughout this paper, stellar masses ($M_\star$) are given in units of $M_\odot$; 
star formation rates (SFRs) are given in units of $M_\odot$~yr$^{-1}$.
The surface mass density (\sigmam) as well as the surface mass density in the central 1~kpc region (\sigmaone) are given in units of $M_\odot$~kpc$^{-2}$; the surface density of SFR (\sigmasfr) as well as that in the central 1~kpc region (\sigmasfrone) are given in units of $M_\odot$~yr$^{-1}$~kpc$^{-2}$.
$L_X$ represents X-ray luminosity at rest-frame 2--10 keV in units of erg s$^{-1}$. 
\lo\ represents the (extinction-corrected) luminosity of the [O III] $\lambda$5007 emission line in units of erg s$^{-1}$.
$L_X$/\mstar\ and \lo/\mstar\ correspond to the X-ray and [O III] luminosity per unit stellar mass, respectively, in units of erg s$^{-1}$ $M_\odot$$^{-1}$.
Reported uncertainties are at the $1\sigma$\ (68\%) confidence level.
A cosmology with $H_0=70$~km~s$^{-1}$~Mpc$^{-1}$, $\Omega_M=0.3$, and $\Omega_{\Lambda}=0.7$ is assumed.

\section{Sample selection and construction} \label{s-sample}

\subsection{The galaxy sample obtained from MaNGA}

The IFU observations from MaNGA map $\gtrsim 10,000$ nearby galaxies at $z = 0.01$--0.15 with 0.5 arcsec $\times$ 0.5 arcsec spaxels \citep{Bundy2015}.
It covers galaxies with a relatively flat distribution in terms of $i$-band absolute magnitude,  
to have spectroscopic coverage out to 1.5 / 2.5 effective radius (\re) for the primary/secondary sample.
There is also a color-enhanced sample to have more objects in the low-density regions of the color-magnitude space.
We start from the \citet{Sanchez2022} MaNGA \texttt{pyPipe3D} value-added catalog, which is built upon reduced data with version 3.1.1 of the MaNGA Data Reduction Pipeline (DRP; \citealt{Law2016}). 
All the galaxies in the main sample can be identified with MANGA\_TARGET1 flag, and there are weights provided by the
DRP catalog to correct the combination of these samples to a volume-limited sample.
We mark broad-line AGNs from \citet{Oh2015, Fu2023} catalogs, which are removed from the sample utilized in this study.\footnote{\citet{Oh2015} selected broad-line AGNs based on the full-width at half-maximum (FWHM) of the \halpha\ emission line of SDSS DR7 spectra. \citet{Fu2023} selected broad-line AGNs based on FWHM of all the Balmer lines utilizing MaNGA spectra directly.}
We also require objects to have $z < 0.1$, so each spaxel spans over $\gtrsim 1$ kpc, which is the minimum scale we are trying to resolve in this study.

\subsection{MaNGA galaxies with X-ray coverage and the selection of X-ray AGNs} \label{ss-mangaxray}

\subsubsection{eROSITA}

We check the MaNGA objects observed by the extended ROentgen survey with an Imaging
Telescope Array (eROSITA) on board the Spektrum-Roentgen-Gamma (SRG) orbital observatory \citep{Predehl2021}, and there are 1870 objects within the coverage.
As MaNGA galaxies are bright foreground galaxies with a field of view of the MaNGA fiber array ranging from 12 to 32'' in diameter, we adopt a 30'' matching radius (comparable to the point spread function size of eROSITA), directly match these galaxies with eROSITA all-sky survey data catalog from the cumulative four scans (eRASS:4), and retain matches within  3$\sigma$ position errors.
Then we adopt the 0.2-2.3 keV flux ML\_FLUX reported in the source catalog,
and convert it to \lx\ assuming galactic absorption with $\Gamma = 1.7$.
For undetected sources, we obtain the background level from the 
background maps and derive the minimum number of counts 
required for a source to be detected in the 0.2--2.3 keV band.
We then derive the corresponding flux sensitivity with the corresponding
energy conversion factor (ECF) = 1.074 $\times$ $10^{-12}$ \citep[e.g.][]{TD2024}, 
Then we convert this flux sensitivity at the galaxy position to \lxlimit\ of the galaxy.

\subsubsection{XMM-Newton} \label{sss-xmm}

With the RapidXMM database \citep{Ruiz2022}, we find 890 MaNGA galaxies within XMM-Newton coverage.
We also directly match those MaNGA galaxies with the 4XMM-DR14 catalog \citep{Webb2020} with a 30'' matching radius,
and adopt sources within the 3$\sigma$ position error.
We convert the X-ray fluxes to \lx\ assuming a power-law model with Galactic absorption and $\Gamma = 1.7$
following the preference order of 4.5--12 keV band, and 0.2--12 keV band, thus minimizing the effects of X-ray obscuration. 
We also estimate \lxlimit\ for the X-ray source to be detected at each MaNGA galaxy position, utilizing the 0.2--12 keV band flux reported in RapidXMM.
A power-law model with Galactic absorption and $\Gamma$ = 1.7 is again assumed through the conversion process.

\subsubsection{Chandra}  \label{sss-chandra}

We make use of the Chandra Source Catalog (CSC) version 2.0 limiting sensitivity properties \citep{Evans2024}, and find 514 galaxies within \chandra\ coverage.
CSC provides 0.5--7 keV band flux sensitivity for a point source to be detected at a given position, 
and we convert this limiting flux to \lxlimit\ with the approach reported in \ref{sss-xmm}.
For sources detected within the 3$\sigma$ \chandra\ position error, we adopt the order of 2--7 keV band, 0.5--7 keV band, and 0.5--2 keV band to estimate \lx, with assumptions similar to those above.

\subsubsection{X-ray AGN selection} \label{ss-xrayagn}

There are 2770 MaNGA galaxies with X-ray coverage from eROSITA, \xmm, or \chandra.
For each galaxy, we estimate the contribution in \lx\ from X-ray binary (XRB), $L_{\rm XRB}$, through a redshift-dependent function
 of $M_\star$ and SFR (model 269, \citealt{Fragos2013}), 
 which is derived utilizing observations in \citet{Lehmer2016}. 
We also estimate the contribution from X-ray hot gas \citep{Mineo2012}, $L_{\rm gas}$.
If \lx\ $>$ 3 $\times$ ($L_{\rm XRB}$ + $L_{\rm gas}$) (which is an empirical selection method for identifying sufficiently excessive X-ray emission from AGNs; see e.g., \citealt{Birchall2022} for details), we classify the object as an X-ray AGN. 
We further classify sources with log \lx\ < 41 as non-AGN even if they meet this criterion, to remove any potential contamination from ultraluminous X-ray sources (ULX).
We note that, among 10 < log \mstar\ < 10.5 galaxies, the detected X-ray sources have luminosities comparable to the expectation from XRB contributions. Thus, we do not select X-ray AGNs with log \mstar\ < 10.5.
For log \mstar\ > 10.5 galaxies, we can safely probe X-ray AGNs with log \lx/\mstar\ $\gtrsim$ 30.5.
For log \mstar\ > 11 galaxies, X-ray AGNs can be probed till log \lx/\mstar\ $\sim$ 30.

We also remove extended X-ray sources in the catalog to avoid any contamination from halo gas.
For sources detected in eROSITA, we adopt the cut of the EXT\_LIKE $>$ 3; for \xmm-detected sources, we adopt the cut of SC\_EXT\_ML > 6; for \chandra-detected sources, we use extent\_flag reported in CSC. Around $20\%$ of the X-ray sources are marked as extended.

\subsection{Optical AGN selection} \label{ss-oagn}

\begin{figure}
\begin{center}
\includegraphics[scale=0.48]{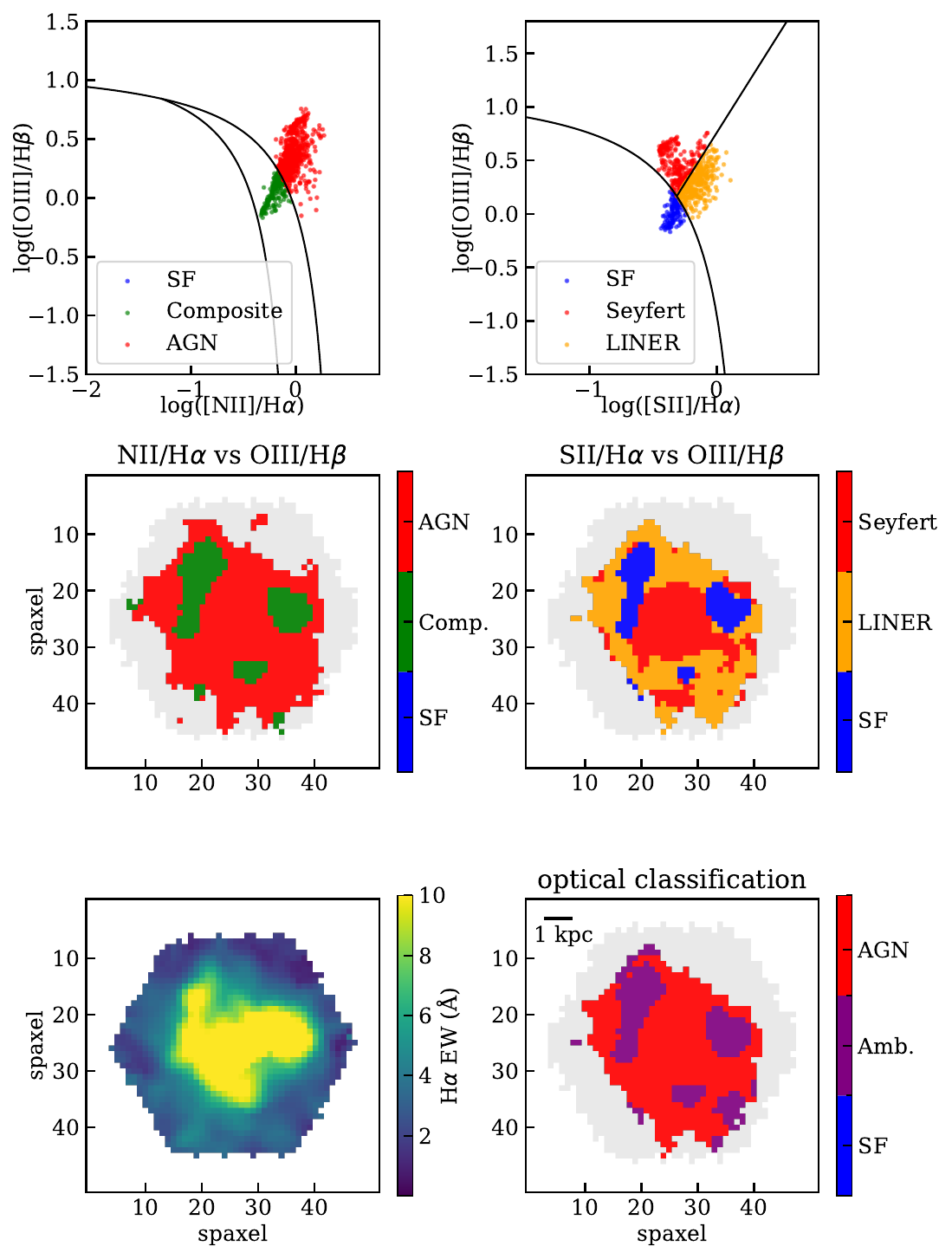}
\caption{Example of an optically-selected AGN. In the top two panels, all the valid spaxels are plotted on the [N II]-based and [S II]-based BPT diagrams. In the middle panels, we show the 2D maps of how these spaxels are classified according to the BPT diagrams. In the lower-left panel, we show the \halpha\ EW map. In the lower-right panel, we show the final spaxel-level classification, which combines all the information.}
\label{opt_agn}
\end{center}
\end{figure}

Among all the MaNGA galaxies with X-ray coverage, we select optical AGNs based on the 
BPT classification diagrams in \citet{Kewley2006}, where different emission line strengths are used to identify the dominant ionization mechanism, as well as the WHAN diagram \citep{Cid2011}, where the equivalent width (EW) of \halpha\ is utilized to remove fake AGNs (post-AGB stars can produce ionized gas with LINER-like line ratios).

To perform the classification, we compute the flux ratios
of [O III]/\hbeta, [N II]/\halpha, and [S II]/\halpha, for valid spaxels -- which are spaxels that have all these emission lines detected at S/N > 3.
We then classify these valid spaxels into classes: AGN, SF, and ambiguous (Amb.).
We classify a spaxel as an AGN spaxel if it is identified as an AGN in both [N II]-based and [S II]-based BPT diagrams, and
has EW(\halpha) $>$ 3~\AA. 
We classify a spaxel as SF if it is classified as SF in both BPT
diagrams. Spaxels that cannot be strictly classified into one of
these categories are classified as ambiguous.
In Figure~\ref{opt_agn}, an example of this spaxel-by-spaxel classification is presented.
A general classification is then obtained by analyzing spaxels within apertures of 1, 2, and 3 kpc in radius, which cover the typical narrow-line region (NLR) size among most AGNs in the MaNGA sample. 
If the number of valid pixels is $> 5$ within a given aperture, and the fraction of AGN spaxels is greater than 20\% of the total and exceeds the SF and ambiguous spaxel fraction, we classify this object as an AGN.
Following the approach in \citet{alban_classifying_2023}, we also compute the average flux ratios of [O III]/\hbeta, [N II]/\halpha, and [S II]/\halpha, as well as the average \halpha\ EW in the galaxy central 1, 2, and 3 kpc apertures, utilizing spaxels where the relevant emission lines have S/N > 3.
We classify an object as an AGN if it is identified as an AGN in both BPT diagrams within any of the given apertures,
and average EW(\halpha) $>$ 3~\AA\ within the aperture is also required.
We then combine the classification results from all the methods: if an object is classified as an AGN by any of the methods, it is recognized as an optically-selected AGN. 

For these optically-selected AGNs, we derive \lo\ based on the [O III] luminosity map and the spaxel-by-spaxel AGN classification map.
To create the [O III] luminosity map, we first derive the uncorrected [O III] luminosity map from the \texttt{pyPipe3D} [O III] flux map. Dust attenuation is then corrected using the Balmer decrement measurements from \halpha\ and \hbeta\ flux maps, assuming the \citet{Calzetti2001} dust extinction law with $R_V = 3.1$, and an intrinsic H$\alpha$/H$\beta$ ratio of 3.1 \citep{Kewley2006}.
We then integrate [O III] luminosity among spaxels that are classified as AGN spaxels as the total \lo.

\subsection{Spatially resolved measurements of star formation} \label{ss-measure-spatial}

To investigate star formation in the central regions, for all the MaNGA galaxies in our sample, we obtain the SFR surface density in the central 1 kpc region, \sigmasfrone, to represent the absolute level of star formation. We also fully leverage the MaNGA IFU data and obtain the radial profile of SFR density (see Section~\ref{sss-measure-sfr}). 
We also combine the \sigmasfr\ map and the \sigmam\ map to calculate the spaxel-by-spaxel deviation from the spatially resolved star formation main sequence (MS) and derive the radial profile of this deviation (see Section~\ref{sss-measure-ms}).
In addition, we produce the map of \dn\ and obtain the spatial profiles, which are closely related to age and can be utilized as another independent measurement of star formation property (see Section~\ref{sss-measure-dn}).

\subsubsection{\sigmasfr\ measurements} \label{sss-measure-sfr}

We derive \sigmasfr\ using the spatially resolved decomposition of the stellar populations 
provided by the \texttt{pyPipe3D} analysis \citep{Sanchez2016,Sanchez2022},
which gives the fractions of luminosity from 273 templates spanning 39 stellar ages (1 Myr –
14.1 Gyr) and 7 metallicities.
We convert the luminosity fractions to stellar mass fractions using the mass-to-light ratios provided
for each Simple Stellar Population (SSP) model.
We then sum up the mass fractions at a certain age (but with different metallicities), 
which will give us the integrated \mstar\ at a certain age for each spaxel.
This will allow us to derive SFR for each spaxel, and the \sfrssp\ adopted throughout this work is the average SFR within $
\sim$100 Myrs, which is calculated by adding all the masses with age $\lesssim$ 100 Myrs and then dividing by the time:

\begin{equation}
{\rm SFR} _{\text{ssp}, t} = \frac{\sum_{age = 0}^{t} M_{\rm \bigstar, age}}{t}
\end{equation}
As stated in \citet{Sanchez2016a}, stellar population parameters obtained from SSP decomposition (e.g., age, metallicity) for MaNGA galaxies typically have uncertainties of 0.1--0.2 dex estimated from Monte Carlo simulations. 
The Monte Carlo uncertainty in \mstar\ derived from SSP decomposition is typically comparable to that of these parameters \citep[e.g.][]{CF2014}.
These small uncertainties provide a reasonable basis for SSP-based SFR measurements, although the precision of such measurements naturally degrades in the regime of very old stellar populations.
We also note that these Monte Carlo uncertainties reflect the statistical errors propagated from the flux measurement uncertainties through the fitting procedure, and do not include systematic effects.
For SFR measurements, a significant portion of the uncertainty arises from the adopted SSP templates, along with other systematic uncertainties related to dust attenuation and star formation history parameterization.
By comparing with SFR values estimated from \halpha\ emission, the scatter of SSP-based SFR values is found to be around 0.3 dex \citep{Sanchez2022}, which demonstrates the general reliability of this type of measurement, considering that SFR averaged over the recent 100 Myr can be very sensitive to star formation history priors and difficult to constrain precisely \citep[e.g.][]{Leja2019}.
We further divide \sfrssp\ per spaxel by the area per spaxel to get \sigmasfr\ measurements.

To obtain \sigmasfrone, we adopt the \sersic\ profiles measured in the NASA-Sloan Atlas (NSA) catalog \citep{Blanton2011} and create elliptical apertures with a major axis of 1 kpc and calculate the mean \sigmasfr\ within the aperture.\footnote{\label{ft-sigmaerror}We also performed Monte Carlo simulations incorporating spectral noise for a set of randomly selected objects to assess the typical uncertainty scale of individual \sigmasfrone\ measurements with \texttt{pyPipe3D}.
The resulting scatter ranges from $\lesssim$ 0.05 dex to $\lesssim$ 0.5 dex.
This scatter can be very small when there is considerable star formation activity, but increases when the central stellar population is older and exhibits little recent star formation, and can be even larger for galaxies with very quiescent centers. \sigmaone, as measured in Section~\ref{sss-measure-ms}, typically shows a scatter of $< 0.1$ dex.
These uncertainties also suggest that systematic uncertainties from, e.g., dust attenuation, star formation history parameterization, and initial mass function constitute a considerable fraction of, or even dominate the uncertainty in \sigmasfrone\ and \sigmaone. The estimated typical uncertainty scales in \sigmasfrone\ and \sigmaone, presented in the right panel of Figure~\ref{sample}, combine the representative Monte Carlo scatters (we take the median of the scatter range) in quadrature with empirical systematic uncertainty floors inferred from the scatter between integrated \texttt{pyPipe3D} estimates and other independent measurements \citep[e.g.][]{Sanchez2022}.}
When measuring the radial profile of \sigmasfr, we also take elliptical apertures with different major axis sizes and use the main \sigmasfr\ of all the spaxels within different elliptical annuli. For all the \sigmasfr\ measurements, including \sigmasfrone\ measurements, we adopt the inclination correction suggested in \citet{Hsieh2017}.

\subsubsection{Deviation from the spatially resolved star formation main sequence} \label{sss-measure-ms}

We obtain \sigmam\ maps utilizing the stellar mass maps provided by the \texttt{pyPipe3D} data products, which are also estimated from SSP decomposition.
By comparing with \mstar\ values estimated from multi-band photometry, the uncertainty of SSP-based \mstar\ measurements is found to be around 0.2 dex \citep{Sanchez2022}, suggesting the general reliability of this type of measurement.
We then measure the \sigmam\ radial profiles as well as \sigmaone\ by taking the mean value of spaxels within the elliptical apertures/annuli with different major axis sizes, with inclination correction performed following \citet{Hsieh2017}.
Following the spatially resolved star formation main sequence obtained in \citet{Hsieh2017}, 
log \sigmasfr\  = 0.715 $\times$ log \sigmam\ $-$ 8.065, we derive the deviation from this MS ($\Delta$MS) for each spaxel, which enables us to calculate the radial profile of $\Delta$MS.

\subsubsection{\dn\ measurements} \label{sss-measure-dn}
The strength of the 4000\,\AA{} break, \dn, is defined as the ratio of the average flux density in the narrow continuum bands 4000--4100\,\AA{} and 3850--3950\,\AA{}. 
\dn\ closely tracks the stellar population age, and can also serve as an indicator for specific SFR (sSFR) at \dn\ $\lesssim$ 1.8 \citep[e.g.][]{Brinchmann2004, Spindler2018}.
To measure spatially resolved \dn\ profiles, we utilize the Data Analysis Pipeline (DAP; \citealt{Westfall2019}) spectral fitting results, which include the best-fitting stellar continuum models without emission line components for all the spaxels.

\subsection{Sample properties}  \label{sss-sampleprop}

\begin{figure*}
   \begin{center}
   \includegraphics[scale=0.55]{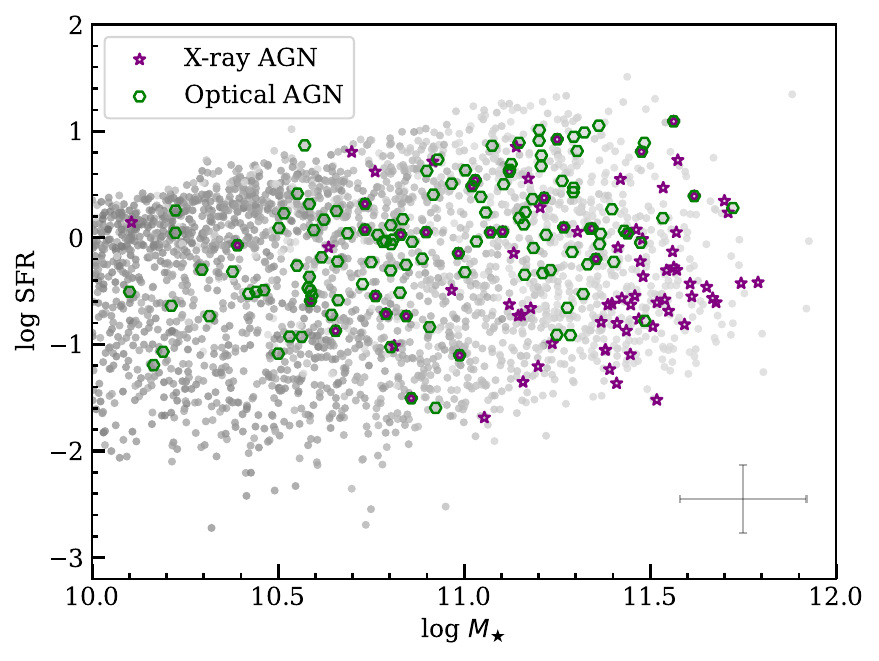}
   ~
   \includegraphics[scale=0.55]{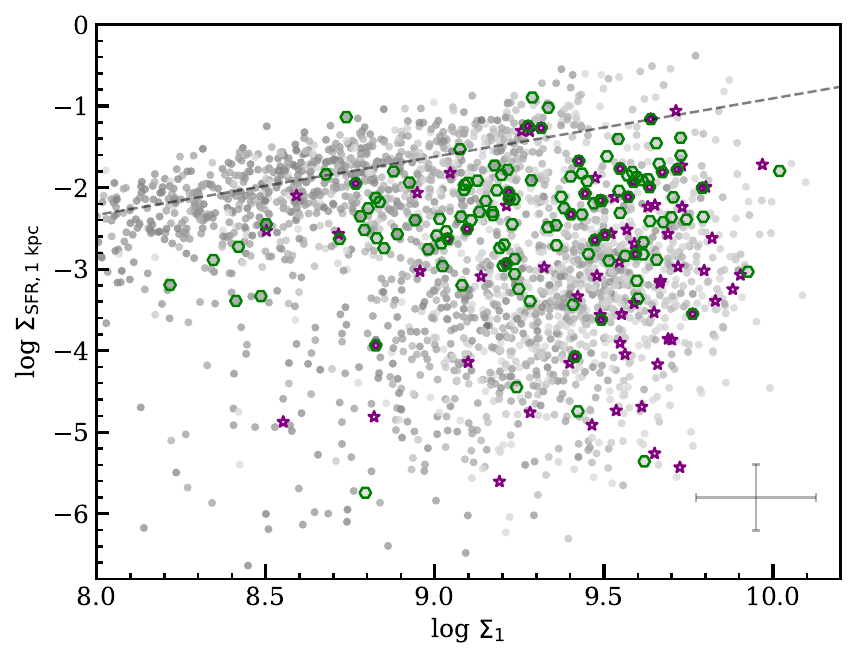}
   \caption{Left: X-ray-selected AGNs and optically-selected AGNs on the \mstar\ vs. SFR plane, as labeled. All galaxies in the sample are displayed as background points, with grayscale intensity encoding their weights.
   The error bar in the corner indicates uncertainties in \mstar\ and SFR derived from \texttt{pyPipe3D} SSP decomposition, estimated by comparison with independent measurements \citep{Sanchez2022}.
   Right: Similar to the left panel, but for the \sigmaone\ vs. \sigmasfrone\ plane. The dashed line depicts the spatially resolved star formation main sequence from \citet{Hsieh2017}. The error bar in the corner indicates approximate uncertainty scales for \sigmaone\ and \sigmasfrone\ (see Footnote~\ref{ft-sigmaerror}).
   }
   \label{sample}
   \end{center}
\end{figure*}

In Figure~\ref{sample}, we plot all the galaxies in the sample on the SFR vs. \mstar\ plane as well as the \sigmasfrone\ vs. \sigmaone\ plane.
X-ray as well as optical AGNs are marked as labeled.
While there is a considerable fraction of AGNs among SF galaxies, we can see that in terms of the spatially resolved star formation main sequence, the central 1 kpc region of most AGNs as well as normal galaxies lie below it (see Section~\ref{ss-compare} for discussions).

\begin{figure}
\begin{center}
\includegraphics[scale=0.6]{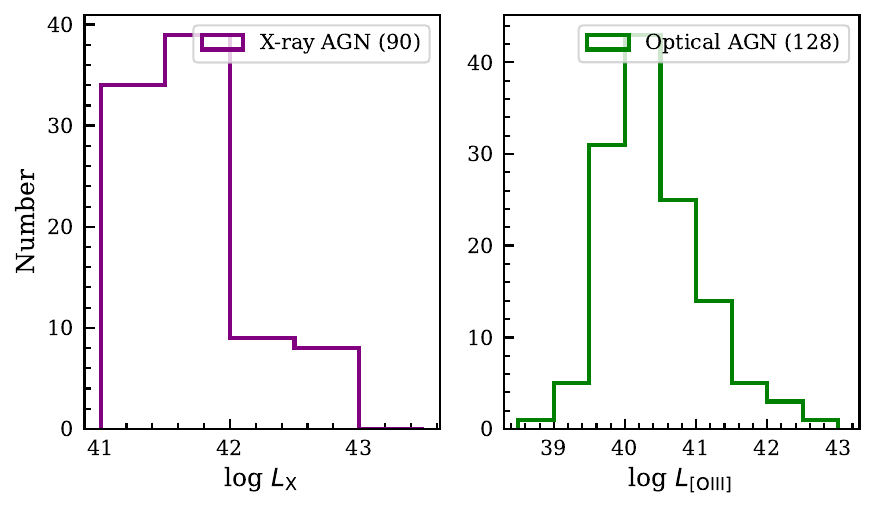}
\caption{Left: The log \lx\ distribution of X-ray AGNs. $L_X$ represents X-ray luminosity at rest-frame 2--10 keV in units of erg~s$^{-1}$. Right: The log \lo\ distribution of optical AGNs. \lo\ represents the (extinction-corrected) luminosity of the [O III] $\lambda$5007 emission line in units of erg~s$^{-1}$.}
\label{agnlum}
\end{center}
\end{figure}

\begin{figure}
\begin{center}
\includegraphics[scale=0.6]{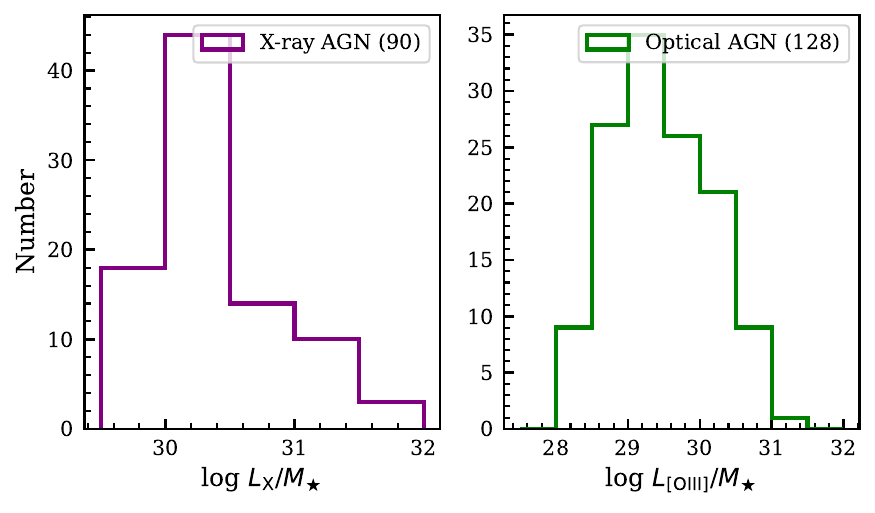}
\caption{Left: The log \lx/\mstar\ distribution of X-ray AGNs.
Right: The log \lo/\mstar\ distribution of optical AGNs.
 $L_X$/\mstar\ and \lo/\mstar\
correspond to the X-ray and [O III] luminosity per unit stellar mass, respectively, in units of erg s$^{-1}$ $M_\odot$$^{-1}$.}
\label{agnlumm}
\end{center}
\end{figure}

In Figure~\ref{agnlum}, we plot the X-ray/[O III] luminosities for X-ray/optical AGNs to approximate the accretion rate of these objects assuming a bolometric correction \citep[e.g.][]{Hopkins2007, Lamastra2009, Duras2020} and a radiative efficiency; in Figure~\ref{agnlumm}, we plot the \lx/\mstar\ and \lo/\mstar\ distributions, which can be used to approximate the specific black hole accretion rate (sBHAR), assuming \mstar\ as a proxy for the BH mass. Most AGNs in our sample have log \lx/\mstar\ $\lesssim$ 31 and log \lo/\mstar\ $\lesssim 30$.
Following the conversion factor between \lx/\mstar\ and Eddington ratio reported in Equation 2 of \citet{Aird2018} (which assumes a constant ratio between \mstar\ and BH mass as well as a constant bolometric correction factor; see \citealt{Aird2018} for details) and the average ratio of \lx\ and \lo\ reported in \citet{Lamastra2009}, these values correspond to very weak accretion activity with Eddington ratio $\lesssim 0.1\%$.

\begin{figure}
\begin{center}
\includegraphics[scale=0.6]{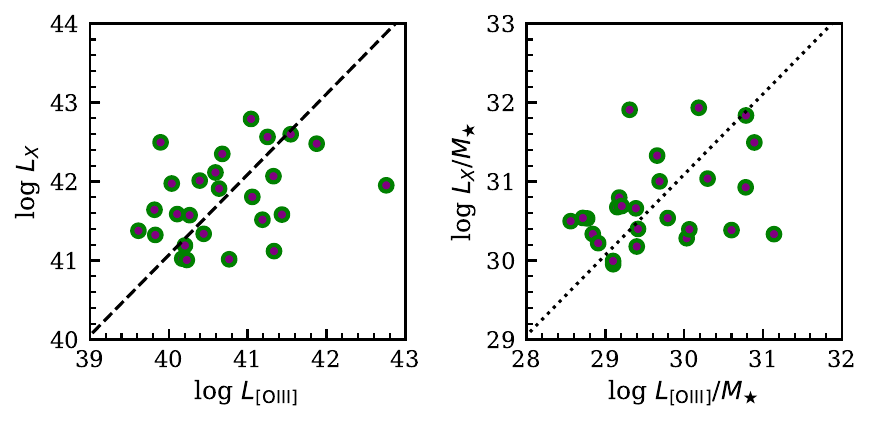}
\caption{Left: \lx\ vs. \lo\ for objects that are identified as both X-ray and optical AGNs. The dashed line is the best-fit log \lx-\lo\ relation reported in \citet{Lamastra2009}. 
Right: \lx/\mstar\ vs. \lo/\mstar\ for objects that are identified as both X-ray and optical AGNs, with the best-fit relation converted from the log \lx-\lo\ relation.
}
\label{xrayvsopt}
\end{center}
\end{figure}

As can be seen in Figure~\ref{xrayvsopt}, log \lx\ and log \lo\, as well as log \lx/\mstar\ and log \lo/\mstar, are loosely correlated among objects that are identified as AGNs in both X-ray and optical bands.
This indicates that the \lx-\lo\ relation is not merely driven by the tendency for more massive galaxies to exhibit higher \lx\ and \lo\ values, and we can compare the sBHAR of X-ray-selected AGNs and optically-selected AGNs through a conversion -- though with caution, as the scatter is significant (the measurement uncertainties of luminosities in this work may contribute substantially to the observed scatter, see Section~\ref{ss-commonfuel} for discussions). 
Throughout the paper, we do the analyses for X-ray AGNs and optical AGNs separately. Since we group X-ray AGNs into two categories with a cut at log \lx/\mstar\ = 30.5, we adopt a log \lo/\mstar\ cut of 29.5 for optical AGNs (which roughly corresponds to log \lx/\mstar\ = 30.5 given the \citet{Lamastra2009} relation represented in Figure~\ref{xrayvsopt}).
We refer to X-ray AGNs with log \lx/\mstar\ > 30.5 as higher-\lx/\mstar\ AGNs, and those with log \lx/\mstar\ < 30.5 as lower-\lx/\mstar\ AGNs. Similarly, we refer to optical AGNs with log \lo/\mstar\ > 29.5 as higher-\lo/\mstar\ AGNs, and those with log \lo/\mstar\ < 29.5 as lower-\lo/\mstar\ AGNs. 
We do not directly refer to these subsamples as higher- or lower-sBHAR X-ray/optical AGNs because of the considerable scatter observed between log \lx/\mstar\ and log \lo/\mstar, which comes from both the measurement uncertainty of these luminosities and the intrinsic uncertainty involved in using \lx/\mstar\ or \lo/\mstar\ as a proxy for the specific accretion rate \citep[e.g.][]{Aird2018, Duras2020, Suh2020}. 
We also caution that most AGNs in the higher-\lx/\mstar\ or \lo/\mstar\ subsamples are still likely to have low Eddington ratios, as discussed in the paragraph above.

\begin{figure*}
\begin{center}
\sidecaption
\includegraphics[width=6 cm]{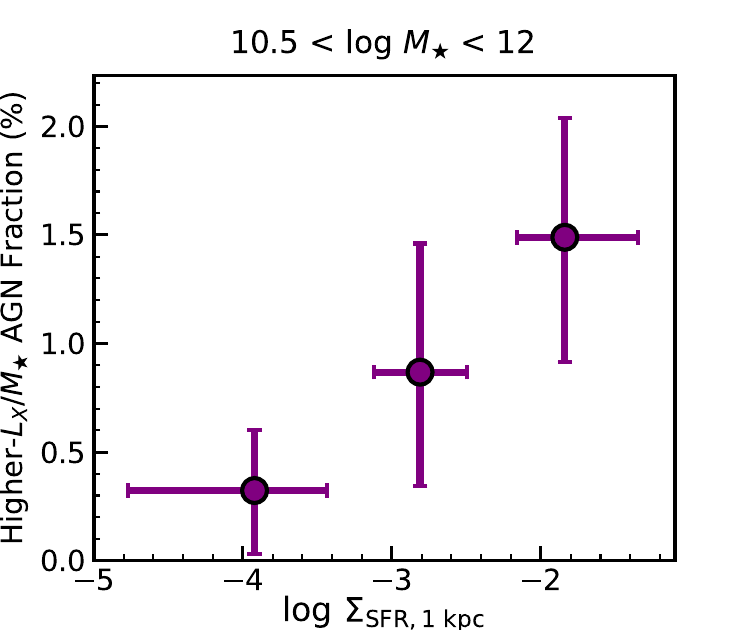}
\includegraphics[width=6 cm]{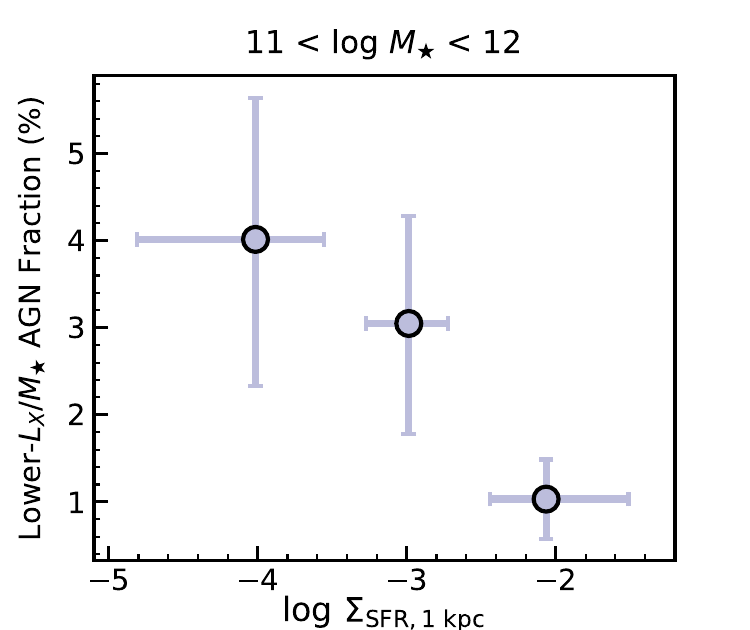}
\caption{Left: The fraction of X-ray AGNs with higher \lx/\mstar\ (log \lx/\mstar\ > 30.5) as a function of \sigmasfrone, among objects with 10.5 < log \mstar\ < 12. 
The y-axis error bars represent the 1$\sigma$ confidence interval of AGN fraction from bootstrapping (i.e. randomly drawing the same number of objects from the sample 1000 times). 
The x-axis error bars represent the 16th and 84th percentiles of the \sigmasfrone\ values. 
Right: Similar to the left panel, but for the fraction of lower-\lx/\mstar\ (log \lx/\mstar\ < 30.5) X-ray AGNs as a function of \sigmasfrone, among objects with 11 < log \mstar\ < 12.}
\label{xrayagn_frac}
\end{center}
\end{figure*}

\begin{figure*}
\begin{center}
\sidecaption
\includegraphics[width=6 cm]{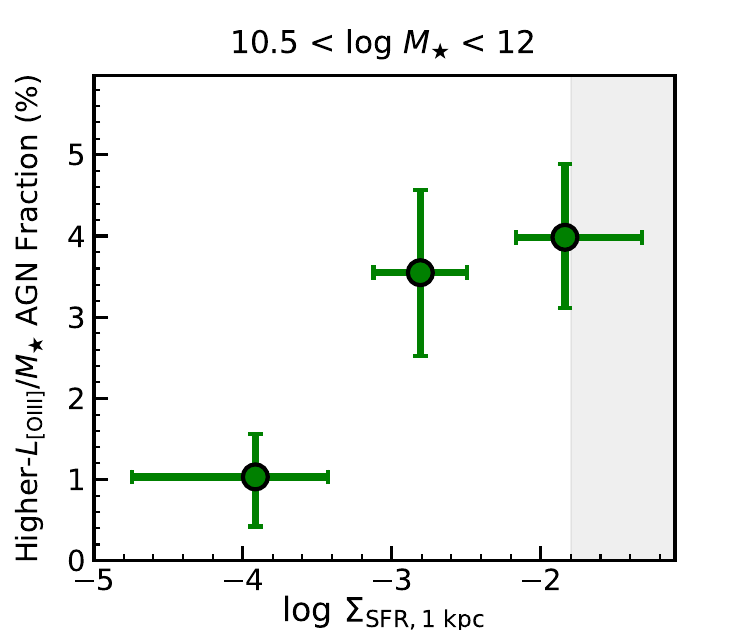}
\includegraphics[width=6 cm]{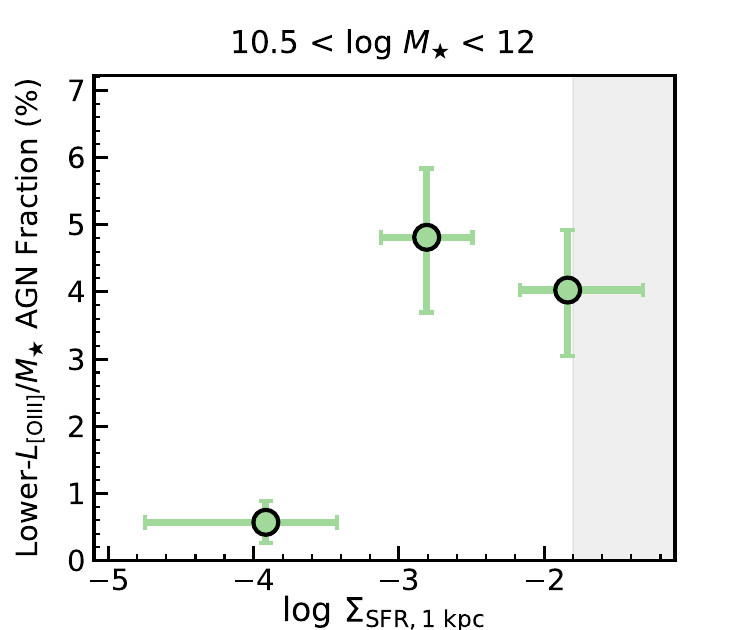}
\caption{Left: The fraction of higher-\lo/\mstar\ (log \lo/\mstar\ > 29.5) optical AGNs as a function of \sigmasfrone, among objects with 10.5 < log \mstar\ < 12. 
The y-axis error bars represent the 1$\sigma$ confidence interval of AGN fraction from bootstrapping.
The x-axis error bars represent the 16th and 84th percentiles of the \sigmasfrone\ values. 
The grey shaded region represents the \sigmasfrone\ range where optical AGN selection is considerably affected by contamination from star formation (see Section~\ref{ss-optsb} for details).
Right: Similar to the left panel, but for the fraction of lower-\lo/\mstar\ (log \lo/\mstar\ < 29.5) optical AGNs as a function of \sigmasfrone\ among 10.5 < log \mstar\ < 12 galaxies.}
\label{optagn_frac}
\end{center}
\end{figure*}

\section{Analyses and results} \label{s-ar}

\subsection{AGN fraction as a function of central SFR density} \label{ss-agnf}

\subsubsection{X-ray-selected AGN fraction} \label{sss-xagnf}

In Figure~\ref{xrayagn_frac}, we show how the X-ray AGN fraction varies as a function of \sigmasfrone.
In order to make sure that the contamination from other X-ray sources will not affect the results (see Section~\ref{ss-xrayagn} for details), we study log \lx/\mstar\ > 30.5 AGNs (i.e., higher-\lx/\mstar\ AGNs) only among objects with log \mstar\ > 10.5 (that have X-ray coverage with \lxlimit/\mstar\ > 30.5 as well), and log \lx/\mstar = 30--30.5 AGNs (i.e., lower-\lx/\mstar\ AGNs) only among galaxies with log \mstar\ $>$ 11 (also with \lxlimit/\mstar\ > 30). 
We divide galaxies satisfying the above criterion into \sigmasfrone\ bins with an equal number of objects per bin, and then calculate the X-ray AGN fraction. When calculating the fraction, we adopt the weight reported in the MaNGA DAP catalog for each object to recover the results for a volume-limited sample.\footnote{We verify that the general trend does not vary qualitatively if no weight is adopted. We also note that since we only use objects with X-ray coverage, the weights to recover a volume-limited sample may not be exactly the same as those provided in the catalog. However, we have verified that the objects with X-ray coverage (as well as those selected using different \lxlimit/\mstar\ thresholds) roughly span the same property space as the full sample.}

We can see that the higher-\lx/\mstar\ AGN fraction (with a median log \lx/\mstar\ $\approx 31$) increases with \sigmasfrone\ in the \mstar\ range we probed.
Due to the limited number of X-ray AGNs with relatively high \lx/\mstar, we are only able to probe them in this large mass range. 
At the same time, among the most massive galaxies, the fraction of lower-\lx/\mstar\ AGNs (with even lower specific accretion rate, log \lx/\mstar\ = 30--30.5) decreases with \sigmasfrone, in contrast to what has been observed for the fraction of higher-\lx/\mstar\ X-ray AGNs among galaxies with relatively smaller \mstar.

\subsubsection{Optically-selected AGN fraction} \label{sss-oagnf}

\begin{figure}
\begin{center}
\includegraphics[width = 4.4cm]{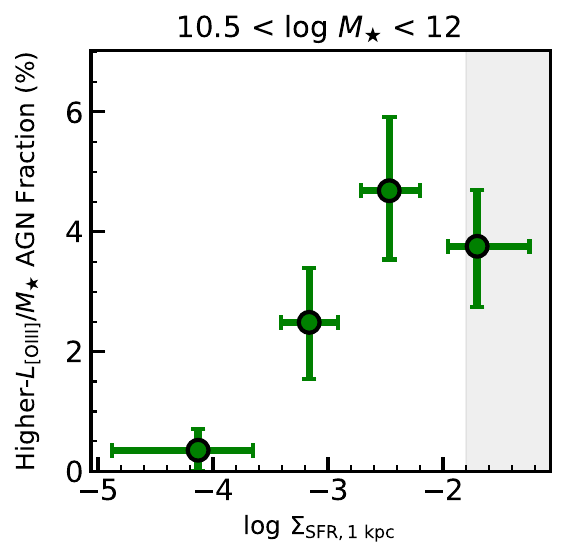}
\includegraphics[width = 4.4cm]{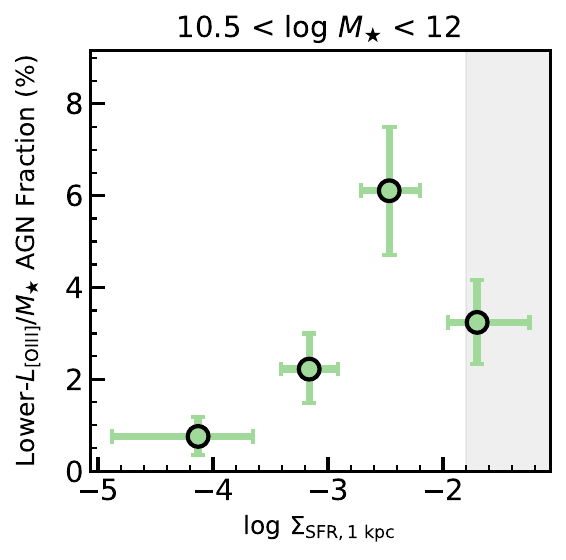}
\caption{Similar to Figure~\ref{optagn_frac}, but presented in four \sigmasfrone\ bins.}
\label{optagn_frac_4bins}
\end{center}
\end{figure}

\begin{figure}
\begin{center}
\includegraphics[scale=0.5]{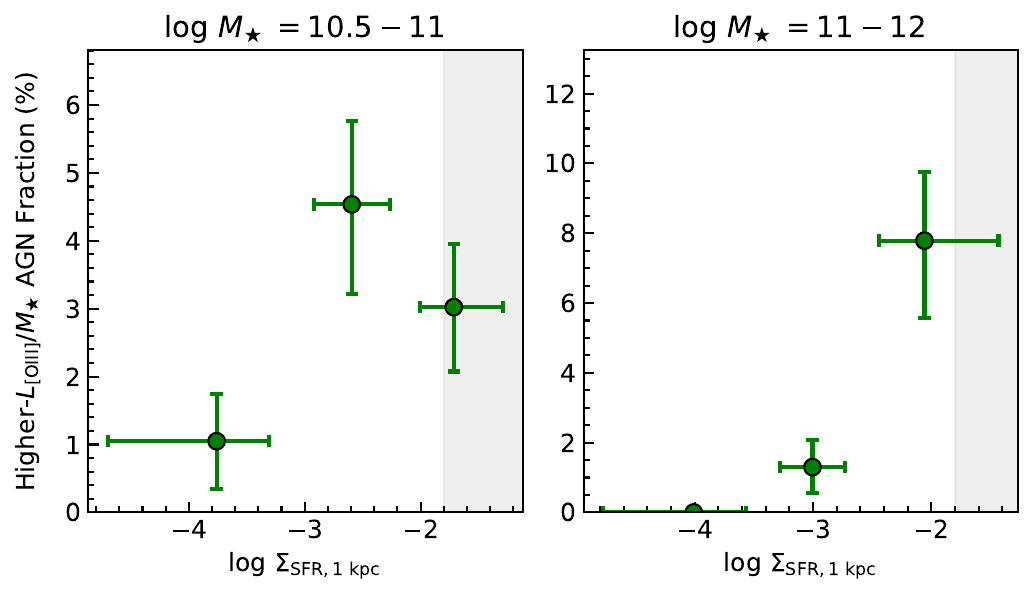}
\caption{Similar to Figure~\ref{optagn_frac}, but for the fraction of higher-\lo/\mstar\ optical AGNs as a function of \sigmasfrone\ among objects with 10.5 < log \mstar\ < 11 (left) and objects with 11 < log \mstar\ < 12 (right). }
\label{optagn_frac_more1}
\end{center}
\end{figure}

\begin{figure}
\begin{center}
\includegraphics[scale=0.5]{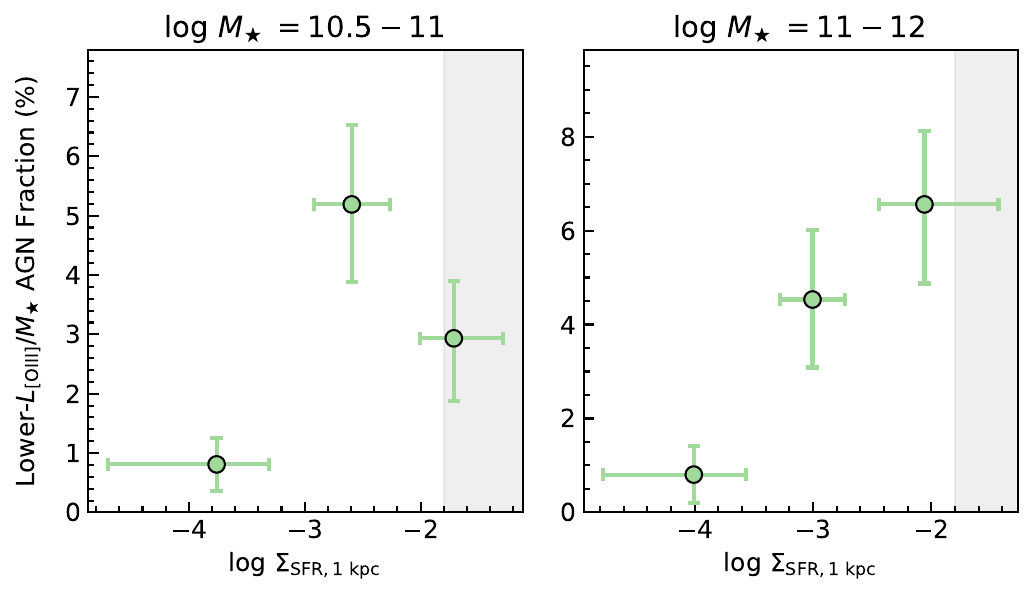}
\caption{The fraction of lower-\lo/\mstar\ optical AGNs as a function of \sigmasfrone, among objects with 10.5 < log \mstar\ < 11 (left) and objects with 11 < log \mstar\ < 12 (right). }
\label{optagn_frac_more2}
\end{center}
\end{figure}

We further plot the fraction of optically-selected AGNs as a function of \sigmasfrone\ among galaxies in our sample.
We study two groups of optically-selected AGNs: those with higher-\lo/\mstar\ values (log \lo/\mstar\ > 29.5) and those with relatively lower-\lo/\mstar\ values (log \lo/\mstar\ < 29.5), to roughly match the \lx/\mstar\ scales in Section~\ref{sss-xagnf} (see Section~\ref{sss-sampleprop} for details).
While it is possible to estimate the [O III] flux upper limit in case of non-detection for each spaxel, since we need to integrate over AGN spaxels to obtain \lo\ so that this would not be possible if the classification itself is not available. 
Furthermore, even when an upper limit of \lo\ can be estimated, proper decomposition is needed to truly estimate the AGN \lo\ upper limit for weak AGNs among SF hosts (see Section~\ref{ss-commonfuel}). Giving an upper limit of \lo\ for an optically-selected AGN to be properly detected for each MaNGA galaxy is beyond the scope of this work, and we left it for future work and instead utilize all the galaxies as the parent sample, similar to previous MaNGA studies of optical AGNs.
For optical AGNs in different \lo/\mstar\ groups, we study them first in the large log \mstar\ $= 10.5$--12 range following the approach in Section~\ref{sss-xagnf} (see Figure~\ref{optagn_frac}), and then in smaller \mstar\ bins to enable better comparison with the X-ray AGN sample (see Figures~\ref{optagn_frac_more1} and \ref{optagn_frac_more2}).

We see that both the higher-\lo/\mstar\ and lower-\lo/\mstar\ optical AGN fractions increase with \sigmasfrone\ when the \sigmasfrone\ is small, but this increasing trend is not significantly present at the high \sigmasfrone\ end, especially for the lower- \lo/\mstar\ AGNs.
When we group galaxies into more \sigmasfrone\ bins (see Figure~\ref{optagn_frac_4bins}), we can clearly observe a decreasing trend at the high \sigmasfrone\ end.
If we study it in smaller \mstar\ ranges (see Figures~\ref{optagn_frac_more1} and \ref{optagn_frac_more2}), we can see that this decreasing trend is mostly associated with the log \mstar\ = 10.5--11 objects. 
Among log \mstar\ = 11--12 galaxies, the optical AGN fraction consistently increases with \sigmasfrone.
While it would be natural to conclude AGN negative feedback at a smaller mass range, we argue that it is actually more complicated, and discuss how selection effects come into play in Section~\ref{ss-optsb}.
In any \mstar\ range or \lo/\mstar\ range, we do not observe elevated optical AGN fraction among low \sigmasfrone\ galaxies as we observed for the lower-\lx/\mstar\ X-ray AGN fraction among massive galaxies (see the right panel of Figure~\ref{xrayagn_frac}), and we discuss this in Section~\ref{ss-optsb} as well.

\subsection{Comparing the star-formation profiles of AGNs and normal galaxies sharing similar global properties} \label{ss-agnprofile}

We also compare the full \sigmasfr\ radial profiles of X-ray/optical AGNs with normal galaxies that share similar global properties.
We have presented how AGN fraction is related to \sigmasfrone, and we note that AGN fraction is related to global SFR as well. 
The full radial profile comparison will help to address whether the link between AGN activity and SFR in the galaxy central region is more than the manifestation of more AGNs among SF galaxies, and why this would be the case.

\subsubsection{Profiles of X-ray-selected AGN hosts} \label{sss-xagnprofile}

Similar to the approach described in Section~\ref{sss-xagnf}, 
we divide the X-ray AGNs into two categories: higher-\lx/\mstar\ (log \lx/\mstar\ > 30.5) AGNs among galaxies with log \mstar\ $=10.5$--12, and lower-\lx/\mstar\ (log \lx/\mstar = 30--30.5) AGNs only among galaxies with log \mstar\ $>$ 11.

\begin{figure*}
\begin{center}
\includegraphics[width = 18cm]{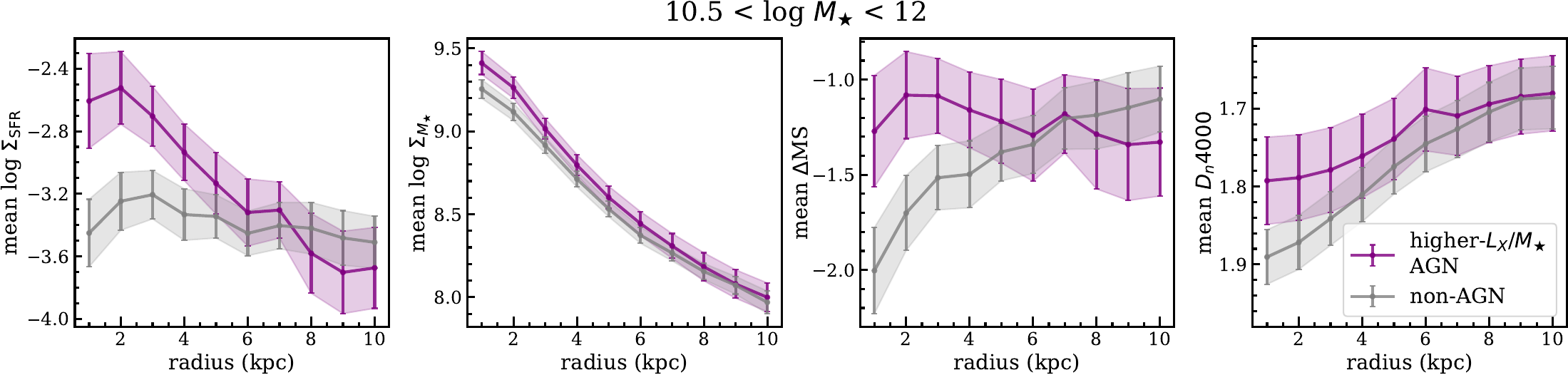}

\vspace{0.3cm}

\includegraphics[width = 18cm]{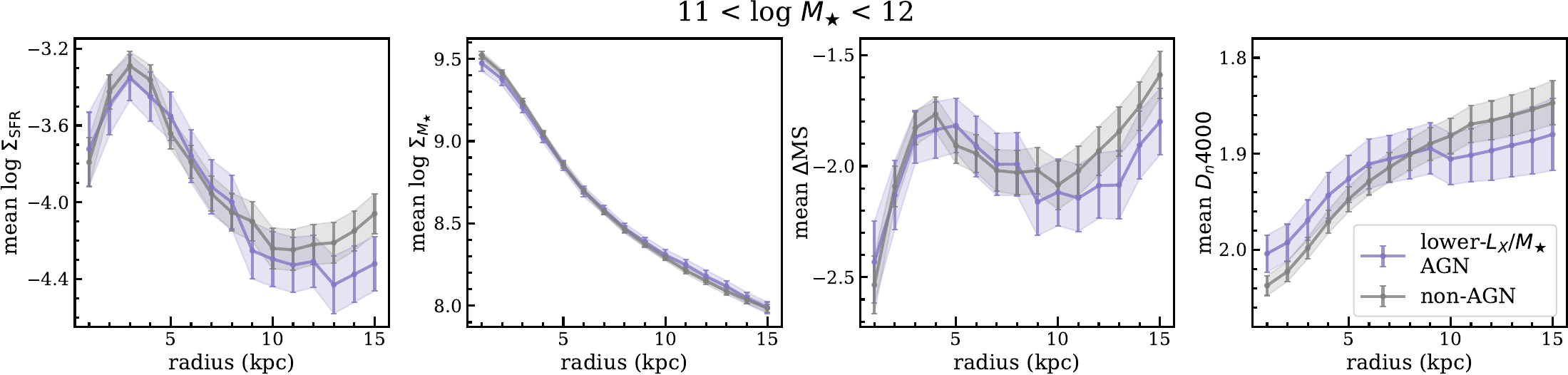}
\caption{Top: The mean \sigmasfr, \sigmam, \dms, \dn\ profiles of higher-\lx/\mstar\ X-ray AGNs compared with normal galaxies in the control sample matched according to \mstar, SFR, and $z$ individually, sampled with a 1 kpc interval in major-axis radius.
Error bars represent the standard error of the mean.
Bottom: Similar to the top panel, but for lower-\lx/\mstar\ X-ray AGNs.
}
\label{xrayagn_profile}
\end{center}
\end{figure*}

When selecting normal galaxies to constitute the control sample, we limit our selection to galaxies with \lxlimit/\mstar\ smaller than the \lx/\mstar\ lower limit of the AGN sample. Furthermore, we exclude optical AGNs to make sure that the contamination in the control sample is as small as possible (we verify that our results do not change if we include optical AGNs in the control sample).
For each X-ray AGN, we select two normal galaxies with the closest \mstar, SFR, and $z$ values utilizing the \texttt{NearestNeighbors} algorithm in the \texttt{scikit-learn} python package.
Then, for both the X-ray AGN sample and the control galaxy sample, we derive the mean radial profile of \sigmasfr, taking elliptical annuli with a width of 1 kpc out to 10/15 kpc (which corresponds to $\sim$ 2 median \re\ for AGNs in the higher-/lower-\lx/\mstar\ sample). 
As can be seen in Figure~\ref{xrayagn_profile}, higher-\lx/\mstar\ AGNs with log \lx/\mstar\ > 30.5 at log \mstar\ $=10.5$--12 show a steeper and more concentrated \sigmasfr\ profile compared to normal galaxies when the total global SFR is roughly the same. In the central regions, higher-\lx/\mstar\ X-ray AGNs have higher \sigmasfr, and the difference in \sigmasfr\ is more prominent when the radius is small. 
In addition to the mean profile of \sigmasfr, we also present the comparison of the mean \sigmam, \dms, and \dn\ profiles.
We note that X-ray AGNs also have slightly higher \sigmam\ in the center -- nevertheless, the differences in \sigmam\ profiles are not as significant as those in the \sigmasfr\ profiles, as a lot of quiescent hosts are included in this study where \sigmam\ cannot trace the gas density very well. 
While both X-ray AGNs and normal galaxies show suppressed star formation relative to the spatially resolved star formation main sequence, as indicated by their mean \dms\ profiles, X-ray AGNs still show a mean \dms\ that is apparently high compared to normal galaxies in the central regions.
In terms of \dn, we can see that X-ray AGNs are generally younger, and the difference is more prominent in the central regions (also indicating high sSFR when \dn\ $\lesssim$ 1.8).
As for the lower-\lx/\mstar\  AGNs (log \lx/\mstar\ = 30--30.5) among log \mstar\ $=11$--12 galaxies, their mean \sigmasfr\ profile, \dms\ profile and \sigmam\ profile do not show noticeable differences when compared with normal galaxies. 
At the same time, we can see that they have slightly younger cores and older outer regions.

\begin{figure*}
\begin{center}
\includegraphics[width = 18cm]{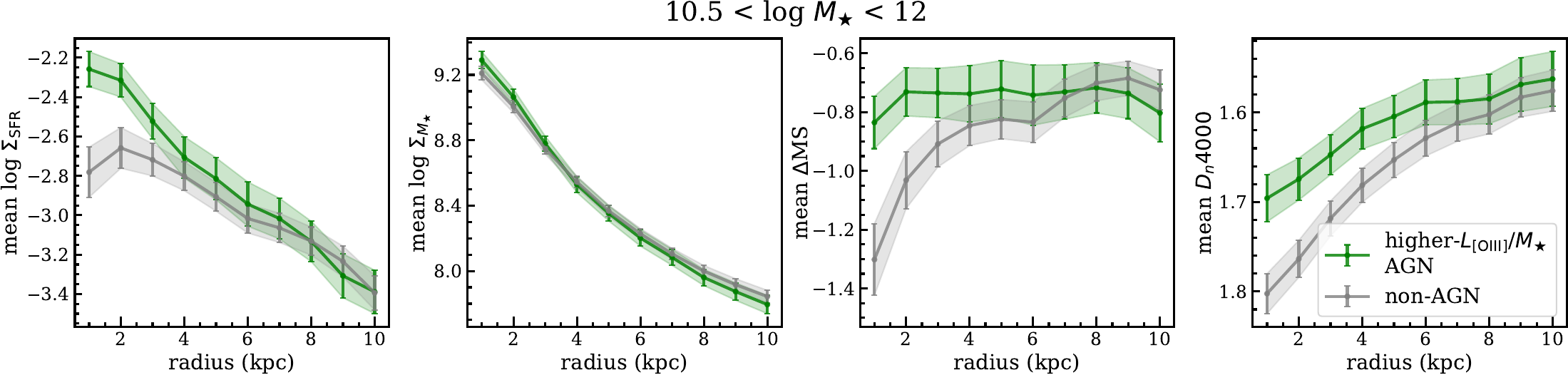}

\vspace{0.3cm}

\includegraphics[width = 18cm]{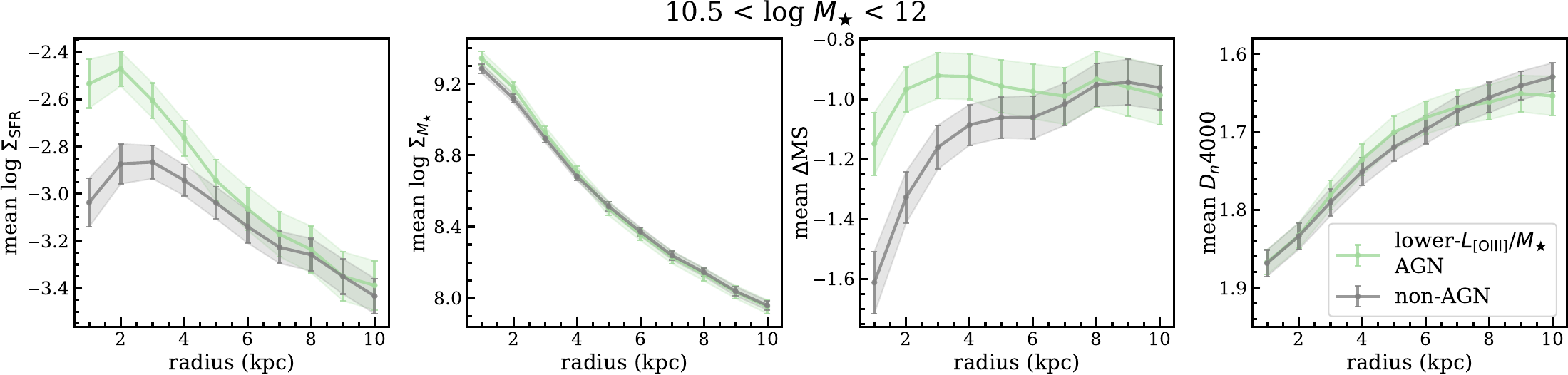}
\caption{Similar to Figure~\ref{xrayagn_profile}, but for higher-\lo/\mstar\ optical AGNs (top) and lower-\lo/\mstar\ optical AGNs (bottom). }
\label{optagn_profile}
\end{center}
\end{figure*}

\subsubsection{Profiles of optically-selected AGN hosts} \label{sss-oagnprofile}
We group the optical AGNs into two \lo/\mstar\ groups as we did in Section~\ref{sss-oagnf}, and compare the profiles of higher- and lower-\lo/\mstar\ optical AGNs with those of normal galaxies following the approach described in Section~\ref{sss-xagnf}.
When constructing the control sample of normal galaxies, we also exclude all the defined X-ray AGNs to make sure that the contamination in the control sample is as small as possible (we verify that our results do not change but become less prominent if we include X-ray AGNs in the control sample). The control sample is again constituted by selecting two normal galaxies with the closest \mstar, SFR, and $z$ values.
As can be seen in Figure~\ref{optagn_profile}, both higher- and lower-\lo/\mstar\ AGNs show higher central \sigmasfr\ compared to normal galaxies. 
They also exhibit higher central \dms\ compared to normal galaxies.
Similar to X-ray AGN hosts, the differences in \sigmam\ profiles of optical AGN hosts compared with those of normal galaxies are not significant.

\section{Discussions} \label{s-dc}
\subsection{Missing optical AGNs among high-\sigmasfrone\ galaxies and low-\sigmasfrone\ galaxies} \label{ss-optsb}

As can be seen in Figures~\ref{optagn_frac} and \ref{optagn_frac_4bins}, the increasing trend of optical AGN fraction with \sigmasfrone\ is not present when the \sigmasfrone\ is high.
It is natural to link this with negative feedback, where AGNs are thought to be responsible for suppressing star formation in galaxy centers. 
However, before we know whether negative feedback is observed or not, we need to account for the bias associated with the BPT-based optical AGN selection method.
Among SF areas, H II regions can produce emission lines that mimic AGN features, and when the AGN is relatively weak or is obscured, the AGN signal may be buried beneath the SF activity: for example, in the BPT diagram we utilized to identify the dominant source of ionization, SF spaxels with weak/obscured AGN signal may fall into the ``ambiguous'' and ``SF'' spaxels.

As suggested by the correlation between SFR and \lo\ reported in \citet{Villa2021}, 
\begin{equation}
{\rm log~SFR = log}~L_{\rm [O III]} - 41.2, 
\end{equation}
higher SFRs require a higher threshold for reliable AGN detection. This also works for the surface density of these values.
We note that $\gtrsim 90\%$ of the optical AGNs in our sample have [O III] surface luminosity density ($\Sigma_{L_{\rm [O III]}}$) in the central 1 kpc region $\gtrsim$$10^{39.7}$ $\rm erg~s^{-1}~kpc^{-2}$.
When $\Sigma_{L_{\rm [O III]}}$ generated by the central star formation is comparable to that generated by the AGN, we consider it difficult to identify the AGN from the BPT diagram. 
This corresponds to log \sigmasfrone\ of $\sim$ $-1.8$.
In reality, we note that AGNs identified via BPT diagrams typically contribute more to the total [O III] flux when they pass the classification line \citep[e.g.][]{Davies2014}.
There are also uncertainties associated with the log SFR-log \lo\ relation: \citet{Villa2021} noted that variations in metallicity and ionization parameter can theoretically lead to a spread of up to 1.1 dex; the empirical scatter of this relation reported ranges from $\approx$ 0.3 to 0.6 dex \citep[e.g.][]{Moustakas2006, Villa2021}.
These indicate that the \sigmasfrone\ threshold for an optical AGN to be very safely detected could be even lower.
Given the contamination threshold, a fraction of optical AGNs may be missed in the highest \sigmasfrone\ bin, so we cannot conclude negative feedback from this.
For galaxies with log \mstar\ $= 10.5$--11, most objects grouped into the highest \sigmasfrone\ bin in Figure~\ref{optagn_frac_more1} have log \sigmasfrone\ from $-2.1$ to $-1.3$. 
Thus, a significant fraction of AGNs in this bin are likely to suffer strongly from selection effects, contributing to the clear drop in the AGN fraction at the high-\sigmasfrone\ end.
For galaxies with log \mstar\ $= 11$--12, most objects grouped into the highest \sigmasfrone\ bin have log \sigmasfrone\ from $-2.5$ to $-1.5$, which has a large fraction of objects likely safe from the selection bias, and the increasing trend can be clearly observed.
While AGNs with higher \lo\ do not naturally translate into higher [O III] surface luminosity density, higher-\lo/\mstar\ AGNs are more likely to have higher $\Sigma_{L_{\rm [O III]}}$. Thus, we can see that the drop at the high-\sigmasfrone\ end is more pronounced among optical AGNs with lower \lo/\mstar.

Also, when we select optical AGNs, we require the emission lines utilized in the study to have S/N $> 3$ and \halpha\ EW > 3\AA. This naturally makes weak/obscured AGNs hard to be valid for classification, especially among quiescent galaxies with low \sigmasfrone\ values.
Thus, for optically-selected AGNs in this study, we are not able to observe the elevated low-\lo/\mstar\ AGN fraction among low-\sigmasfrone\ galaxies as we observed for X-ray-selected AGNs, leading to the observed discrepancy.

\begin{figure*} 
\begin{center}
\sidecaption
\includegraphics[width = 6 cm]{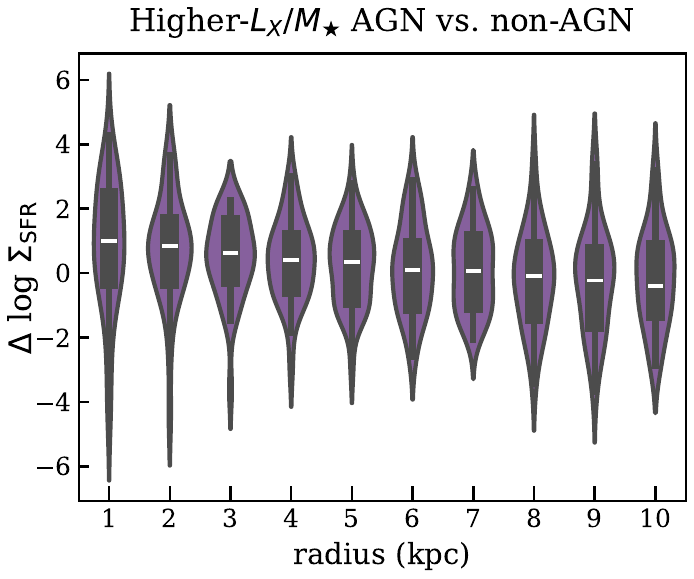}~
\includegraphics[width = 6 cm]{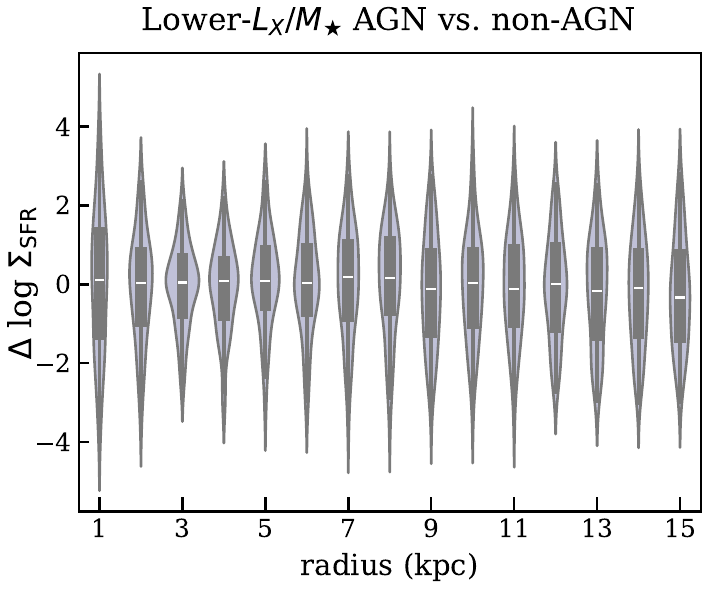}
\caption{Left: Violin plot showing the distribution of differences in \sigmasfr\ between higher-\lx/\mstar\ AGNs and matched control galaxies at each radius. 
Similar to the mean \sigmasfr\ profiles presented in Figure~\ref{xrayagn_profile}, clear differences in the medians are also observed, especially in the central regions. However, the distributions of $\Delta$ log \sigmasfr\ for individual matched pairs remain broad, spanning both positive and negative values.
Right: Similar to the left panel, but for lower-\lx/\mstar\ AGNs.
The $\Delta$~log~\sigmasfr\ distributions are also broad, while no clear differences in the medians are observed in galaxy central regions (aligning with the lack of significant differences in the mean \sigmasfr\ profiles).
}
\label{xrayagn_diffprofile}
\end{center}
\end{figure*}

\begin{figure*}
\begin{center}
\sidecaption
\includegraphics[width = 6 cm]{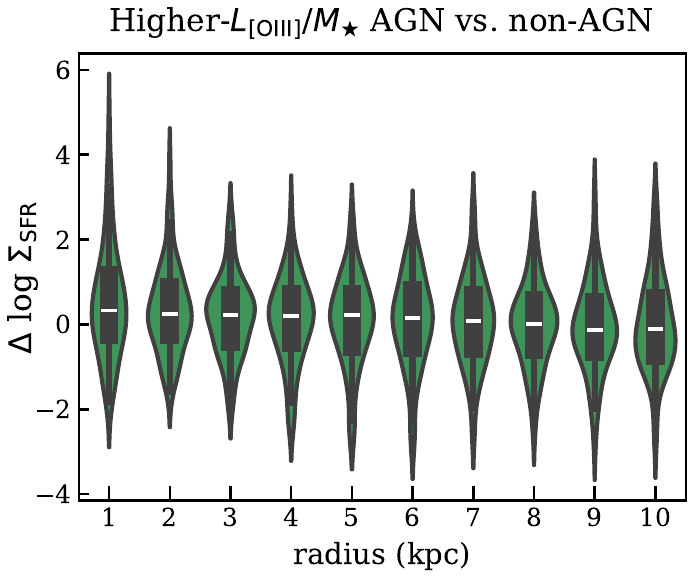}~
\includegraphics[width = 6 cm]{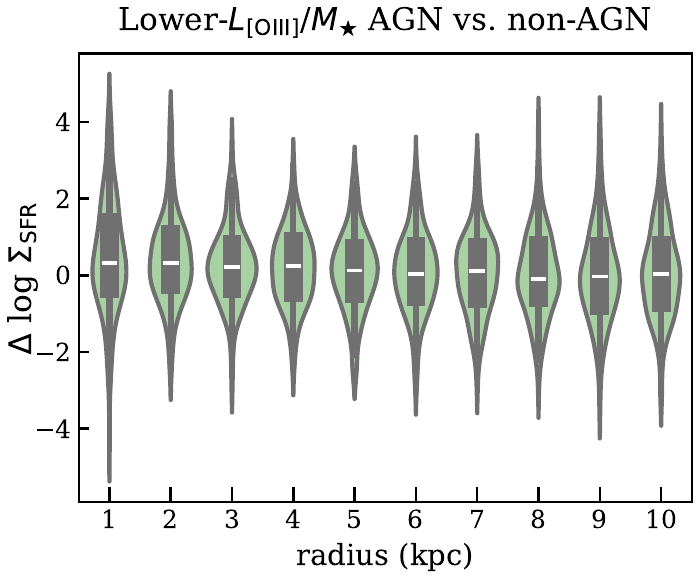}
\caption{Similar to Figure~\ref{xrayagn_diffprofile}, but for higher-\lo/\mstar\ optical AGNs (left) and lower-\lo/\mstar\ optical AGNs (right).
Consistent with the clear differences in the mean \sigmasfr\ profiles presented in Figure 12, offsets in the median values of the distributions (indicated by short horizontal lines) are observed for both groups in the central regions of galaxies.
The overall distributions of $\Delta$ log \sigmasfr\ for individual matched pairs are broad, extending significantly into both positive and negative values.
}
\label{optagn_diffprofile}
\end{center}
\end{figure*}

\subsection{Is coeval growth in tension with AGN feedback?} \label{ss-commonfuel}

Unlike optically-selected AGNs, the fraction of higher-\lx/\mstar\ X-ray AGNs with log \lx/\mstar\ $> 30.5$ shows a steadily increasing trend with \sigmasfrone\ (see the left panel of Figure~\ref{xrayagn_frac}), 
consistent with the expectation from the coeval growth picture arising from the common origin of gas in the vicinity of the BH and in the central region of galaxies. 
We note that X-ray selection is not completely safe from missing obscured AGNs, especially since a fraction of sources are selected in the eROSITA sky coverage, which is mainly sensitive in the soft band rather than the hard band (the obscuration effect is minimized in the hard band). If we consider the obscuration level of AGN structure itself (e.g., torus) does not vary significantly with \sigmasfrone, the obscuration level from the host galaxy component (which also leads to considerable obscuration; e.g., \citealt{goulding2018, gilli2022}) could increase with \sigmasfrone. As the increasing trend is present even with potential obscuration bias, we consider the picture of common fuel and coeval growth to be stable. 
If we look at the \dn\ profiles (see the top panel of Figure~\ref{xrayagn_profile}), we can see that these higher-\lx/\mstar\ X-ray AGNs, as well as all the optical AGNs with higher \lo/\mstar\ (see Figure~\ref{optagn_profile}), generally have younger stellar populations compared to galaxies with similar \mstar\ and SFR, suggesting different star formation histories -- galaxies that host these AGNs are more likely to have more recent star formation episodes so that the average age appears lower, and the cold gas replenishment also triggers AGN activity.

Due to the limited sample size, we cannot statistically prove how significant \sigmasfrone\ is as an indicator for AGN activity or BH growth compared to total SFR. However, analysis results in Figures~\ref{xrayagn_profile} and \ref{optagn_profile} already demonstrated that the link between AGN activity and \sigmasfrone\ is more than manifesting the link between AGN and global SFR, as these AGNs and control galaxies have similar global SFRs.
In \citet{Ni2021}, we found that the projected central surface mass density within 1 kpc, \sigmaone, is more effective in predicting the BH growth level compared to other compactness parameters among SF galaxies. Now we know that this can be understood at a phenomenological level by the slope of the \sigmasfr\ radial profiles -- the differences between AGNs and normal galaxies in \sigmasfr, thus available gas fuel, are more prominent in the central regions.

However, there is well-established observational evidence for AGN evacuating gas in the central parts of galaxies \citep[e.g.][]{Burillo2014,Fluetsch2019, Parlanti2025}, so one will naturally expect reduced \sigmasfrone\ in these cases -- how can these results be consistent with more AGNs among galaxies with a higher level of central star formation?

In Figure~\ref{xrayagn_diffprofile}, instead of comparing the mean \sigmasfr\ as a function of major axis radius of X-ray AGNs and matched normal galaxies, we show the distribution of the difference in log \sigmasfr\ ($\Delta$ log \sigmasfr; log \sigmasfr (AGN) $-$ log \sigmasfr (control galaxy) at the individual level) at different radii.
Similar to the obvious difference in mean log \sigmasfr\ as shown in Figure~\ref{xrayagn_profile}, the median log \sigmasfr\ of X-ray AGNs is also higher in the central regions.
At the same time, the wide distribution of $\Delta$log \sigmasfr\ suggests the unreliability of deriving a universal conclusion for the AGN feedback scenario via a limited number of objects, particularly a single observation (which has been a widely adopted approach), as it will easily lead to a conclusion of positive feedback or negative feedback -- which could be in action for this particular object, but not all the others.
When considering the overall trend, while we attribute the link between AGN and higher central \sigmasfr\ (on average) to coeval growth from a common fuel supply, the wide distribution of $\Delta$ log \sigmasfr\ can be attributed to various factors.

We first need to account for AGN variability.
The X-ray luminosity of AGNs can vary by orders of magnitude over a Myr timescale \citep[e.g.][]{Hickox2014}, and the SFR we adopted in this study is measured over even longer timescales. The observed X-ray AGNs with a certain \lx/\mstar\ criterion might not fulfill this criterion over most of the time where SFR is estimated; similarly, the normal galaxy used for comparison might be an X-ray AGN in recent history.
While \lo\ also exhibits variability and is subject to the same timescale effect mentioned above, compared with the X-ray emission, [O III] emission is much closer to a long-term average measurement,
as it comes from the narrow-line region that is further away from the central SMBH (at the scale of $\sim$ 10$^4$--10$^5$ light-years) and has an extended spatial distribution.
In this case, one would expect smaller scatters of the $\Delta$log \sigmasfr\ compared to normal galaxies, if the variability is playing a smaller role here.
We do observe that the scatter is slightly smaller, particularly at the central radii.
However, the scatter is still considerable, which might relate to the optical AGN selection bias in this study (see Section~\ref{ss-optsb}), so that there are still unclassified AGNs mixed in the matched control galaxy sample.

In addition to all these, there might be real hints of feedback in action.
It has been argued that negative feedback on star formation is most efficient in powerful AGNs \citep[e.g.][]{Cicone2014}, as the powerful AGN winds could blow out gas in the central regions and stop the gas from fragmentation \citep[e.g.][]{Zubovas2017}.
For AGNs that are not especially luminous, such as objects in our sample, small-scale outflows can inject turbulence in the ISM, and/or heat the gas, thus also weakening the star formation, especially in the central region \citep[e.g.][]{Parlanti2025}. 
At the same time, positive AGN feedback is also predicted and observed by compressing the ISM and enhancing the SFR \citep[e.g.][]{Maiolino2017, Shin2019}.
We cannot confidently argue that we have observed strong evidence for positive AGN feedback, as the overall elevated \sigmasfr\ spans areas larger than what AGNs at this low level of luminosity can impact.
However, we might have observed evidence for negative AGN feedback, though very slight, as the \sigmasfr\ profiles in Figures~\ref{xrayagn_profile} and \ref{optagn_profile} either show a small decline in the central $\sim$ kpc region (though this decline is small compared to the absolute value of \sigmasfr), or show a flattening of the inwardly increasing trend within the central $\sim$ kpc.
Under the negative AGN feedback scenario, one might expect more extended tails toward negative $\Delta$log \sigmasfr\ and a smaller median $\Delta$log \sigmasfr\ value in the central region among higher-\lo/\mstar\ optical AGNs compared to lower-\lo/\mstar\ optical AGNs in Figure~\ref{optagn_diffprofile}, if more powerful AGNs launch more powerful outflows. 
We observe slightly more extended negative $\Delta$log \sigmasfr\ tails over the central region among lower-\lo/\mstar\ optical AGN, but we note that the selection effect is also playing a role in the optical AGN identification as discussed earlier, and the differences are not significant, so we cannot reliably test the hypothesis that central star formation suppression is more pronounced or not among AGN hosts with higher AGN luminosities from these plots. 
Also, we note that \lo\ obtained from summing over the [O III] flux reported in \texttt{pyPipe3D} data products in AGN-dominated spaxels may not be a robust indicator of the true AGN luminosity.  As we have discussed, weak/obscured AGN emission is hard to distinguish from the emission lines from SF activity; also, AGN outflows can broaden the [O III] emission line, and the shifted wings may contribute significantly but are not accounted correctly given the current \texttt{pyPipe3D} line-measurement method (a single Gaussian fitting; \citealt{Sanchez2016}).
Thus, our subsamples divided according to \lo/\mstar\ might not distinguish true specific accretion power very clearly, which is also shown in Figure~\ref{agnlumm} -- a better measurement of the \lo\ from the AGN will help to further investigate the negative AGN feedback scenario, if objects with higher- and lower-sBHAR are better separated.

\begin{figure}
\begin{center}
\includegraphics[width = 7 cm]{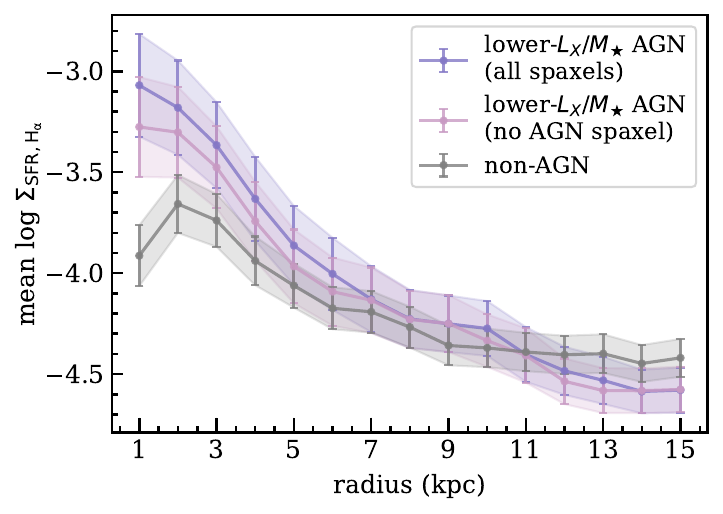}
\caption{The SFR surface density profiles estimated from H$\alpha$, of X-ray AGNs with lower \lx/\mstar\ (log \lx/\mstar\ < 30.5) and normal galaxies utilized in Figure~\ref{xrayagn_profile}. We can see that even the mean \sigmasfr\ estimated from H$\alpha$ emission from spaxels that are not AGN-dominated is elevated compared to that of non-AGNs in the control galaxy sample, suggesting that a minute level of SF enhancement is likely among the hosts of these very weak AGNs, when we select an SF indicator that is sensitive to very recent SF activity.}
\label{lowlagn_sfrha}
\end{center}
\end{figure}

Among massive galaxies with log \mstar\ $> 11$, we observe a decreasing trend of notably low sBHAR (log \lx/\mstar\ < 30.5) AGNs with increasing \sigmasfrone\ (see the right panel of Figure~\ref{xrayagn_frac}). We interpret this as being associated with both large \mstar\ and the low sBHAR, where stochastic fuelling from condensed hot halo gas serves as the major fuel. Hot halo gas fuelling has long been proposed as a source for powering low-sBHAR AGNs, and both the trend of increasing low-sBHAR radio AGN fraction and X-ray AGN fraction with \mstar\ has been observed among massive galaxies \citep[e.g.][]{BH2012, Ni2023}.
Among massive galaxies with log \mstar\ $> 11$, lower \sigmasfrone\ is indeed associated with galaxies with higher \mstar\ and older stellar populations. These galaxies tend to reside in larger halos \citep[e.g.][]{WT2018}, which also contain more fuel to power AGNs with low levels of accretion activity through stochastic condensation of halo gas, so that an elevated lower-\lx/\mstar\ AGN fraction is observed at the low-\sigmasfrone\ end. For galaxies with similar \mstar, older galaxies tend to form earlier in more massive halos \citep[e.g.][]{Rodriguez2015, WT2018} that have more fuel for the central BH -- this explains why log \lx/\mstar\ < 30.5 AGN hosts on average have older stellar populations in the outskirts compared to matched normal galaxies (see the bottom panel of Figure~\ref{xrayagn_profile}), which is consistent with their formation history, but also younger cores, as a result of slightly enhanced star formation level in the center from a larger reservoir of condensed halo gas. 
We also do not observe an obvious difference in the mean \sigmasfr\ profiles of these X-ray AGNs and normal galaxies (see Figure~\ref{xrayagn_profile}). 
We note that SFR in this study is derived based on decomposing the stellar continuum (see Section~\ref{sss-measure-sfr}), which is not very sensitive to a small level of very recent star formation (at the $\sim$10 Myr scale) when the overall stellar population is old. \sigmasfr\ derived from SSP decomposition is more stable for the average SFR over a longer period, e.g., the 100 Myr scale used in this study. Higher levels of H$\alpha$ emission can indeed be observed among these massive AGN hosts that cannot fully be attributed to AGN emission, indicating very weak star formation activity (which supports that what fuels the AGN must fuel the galaxy at the same time, to some extent) that becomes hard to detect when averaged out to longer timescales (see Figure~\ref{lowlagn_sfrha}).\footnote{We estimated the H$\alpha$-based \sigmasfr\ following the method in Section 3.1 of \citet{Spindler2018}, where the relation from \citet{Kennicutt1998} for a \citet{Salpeter1955} initial mass function is adopted. We note that there are more caveats compared to the advantages of estimating SFR from $ H\alpha$. Due to the complexity in properly decomposing the contribution from AGN and star formation in the H$\alpha$ emission line, which is beyond the scope of this work, we do not adopt H$\alpha$-based \sigmasfr\ measurements in this study.} 
In general, we observe more low-\lx/\mstar\ AGNs among galaxies with low \sigmasfrone, which depleted their cold gas earlier but have more hot halo gas fuel available.
Also, compared to the mean \sigmasfr\ profiles of higher-\lx/\mstar\ AGNs among less massive galaxies, the mean \sigmasfr\ profile of massive log \lx/\mstar\ < 30.5 AGN hosts shows a larger decline in the central $\lesssim$ 3 kpc region.
For low-accretion-rate AGNs among massive galaxies, jet activity is considered to be very prevalent \citep[e.g.][]{Best2005}.
AGN jets are known to be a key mechanism for maintenance-mode AGN feedback, during which the surrounding gas is heated and disturbed, and cannot cool and form stars efficiently. Simulation results \citep[e.g.][]{Gaibler2012} predict that jets can create a small cavity in the galaxy center ($\lesssim 3$ kpc), and increase SFR in the outer regions --  these effects might explain the central \sigmasfr\ decline we observed among massive log \lx/\mstar\ < 30.5 AGN hosts that is more noticeable (in both scale and amplitude) compared to that among other AGN hosts.
While a single AGN active episode is short, the cavities left behind can persist on timescales comparable to those over which the SSP-based SFR is estimated ($\sim$ 100 Myr). 
In the bottom panel of Figure~\ref{xrayagn_profile}, we see the central \sigmasfr\ decline almost equally in terms of AGN and non-AGN, which would be consistent with the effects of AGN jet-mode feedback on the relevant timescales.

\subsection{Comparison with previous studies} \label{ss-compare}

\begin{figure*}
\begin{center}
\includegraphics[scale=0.41]{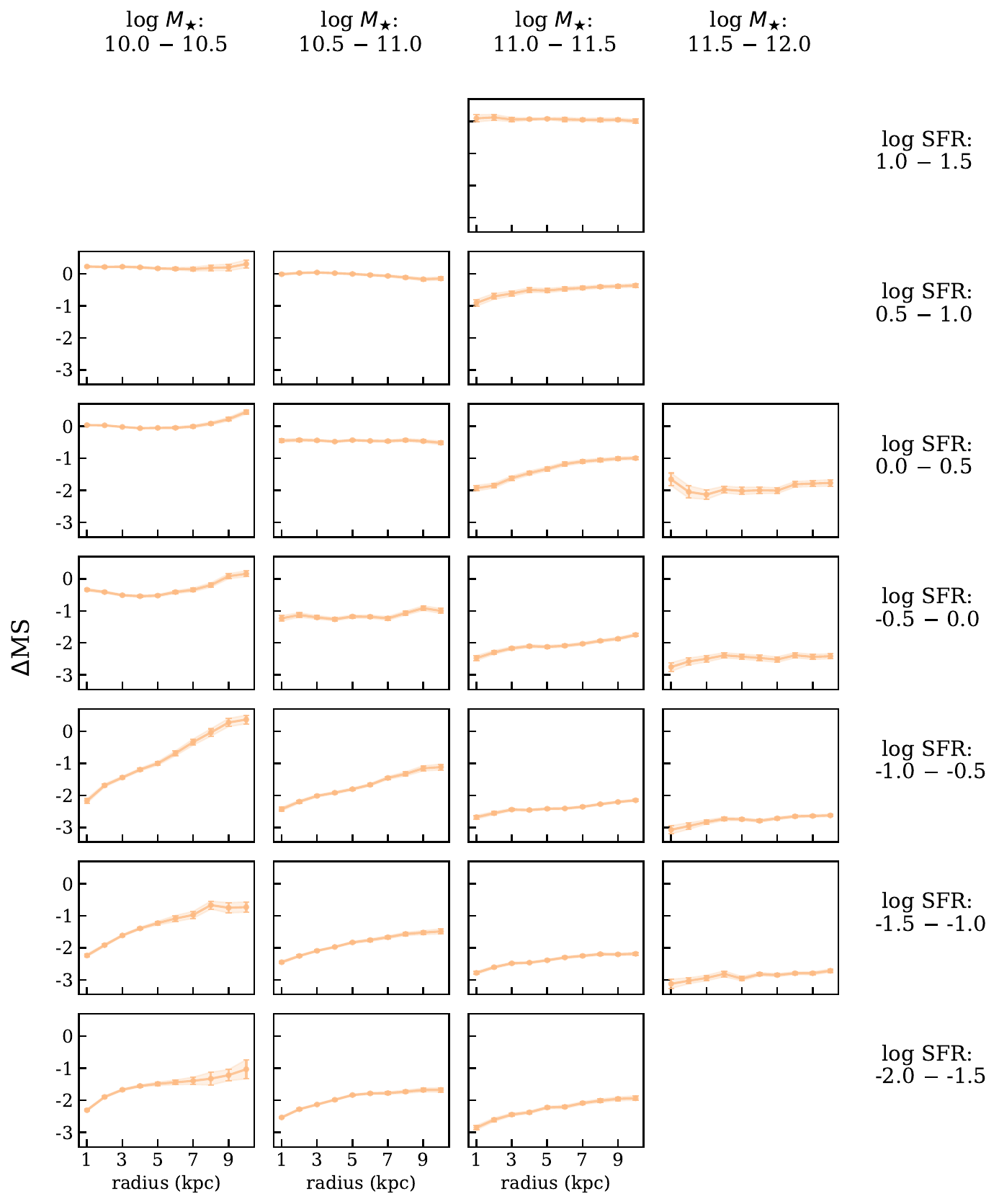}
\includegraphics[scale=0.41]{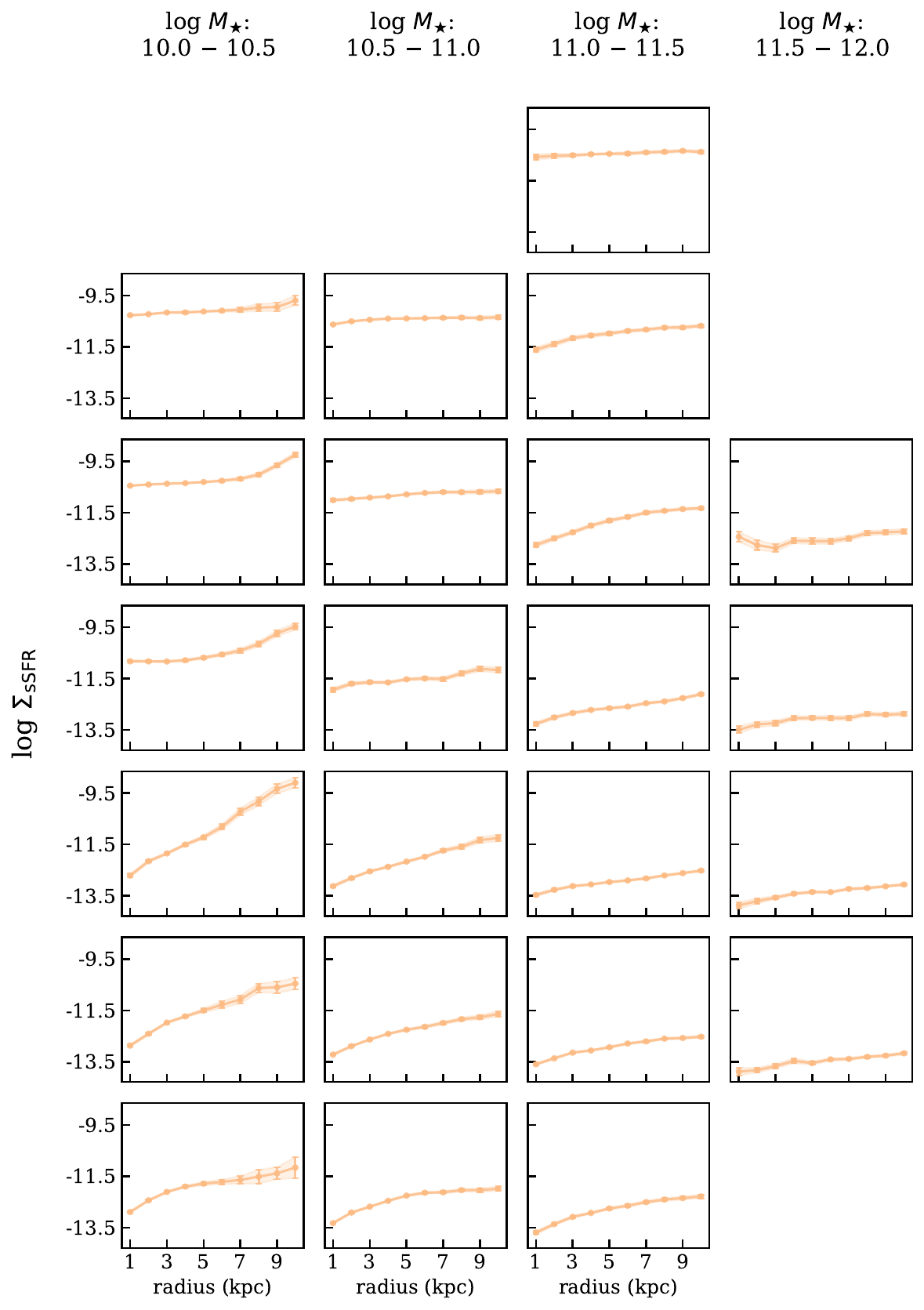}
\caption{Left: Mean \dms\ profiles of all the galaxies in the sample grouped according to their \mstar\ and SFR values. Error bars represent the standard
error of the mean. Right: Similar to the left panel, but for \sigmassfr\ profiles.}
\label{all_deltamsprofile}
\end{center}
\end{figure*}

In Figure~\ref{all_deltamsprofile}, we plot the mean \dms\ profiles as well as the surface density profiles of specific SFR (\sigmassfr) of all the galaxies in the sample grouped into different \mstar\ and SFR bins. 
We can see that galaxies in the MaNGA sample have suppressed central star formation properties on average, while most of them do not host detectable AGNs.\footnote{In general, log \sigmassfr\ profiles are similar to \dms\ profiles. Since galaxies have higher \sigmam\ in the central regions, in some cases, one can observe 
 a slightly decreasing \sigmassfr\ profile towards the center, but a relatively flat \dms\ profile, as the slope of the spatially resolved star formation main sequence is smaller than one.}
Thus, it is not surprising to observe suppressed SFR among AGN hosts \citep[e.g.][]{minijpas2024} or low gas fraction among AGN regions \citep[e.g.][]{Ellison2021} -- galaxies on average have suppressed centers at low redshift, and AGN regions are just located in the centers.
A causal relation between AGN feedback and the suppression of central SFR concluded this way needs further caution.

\begin{figure*}
\begin{center}
\sidecaption
\includegraphics[width = 12 cm]{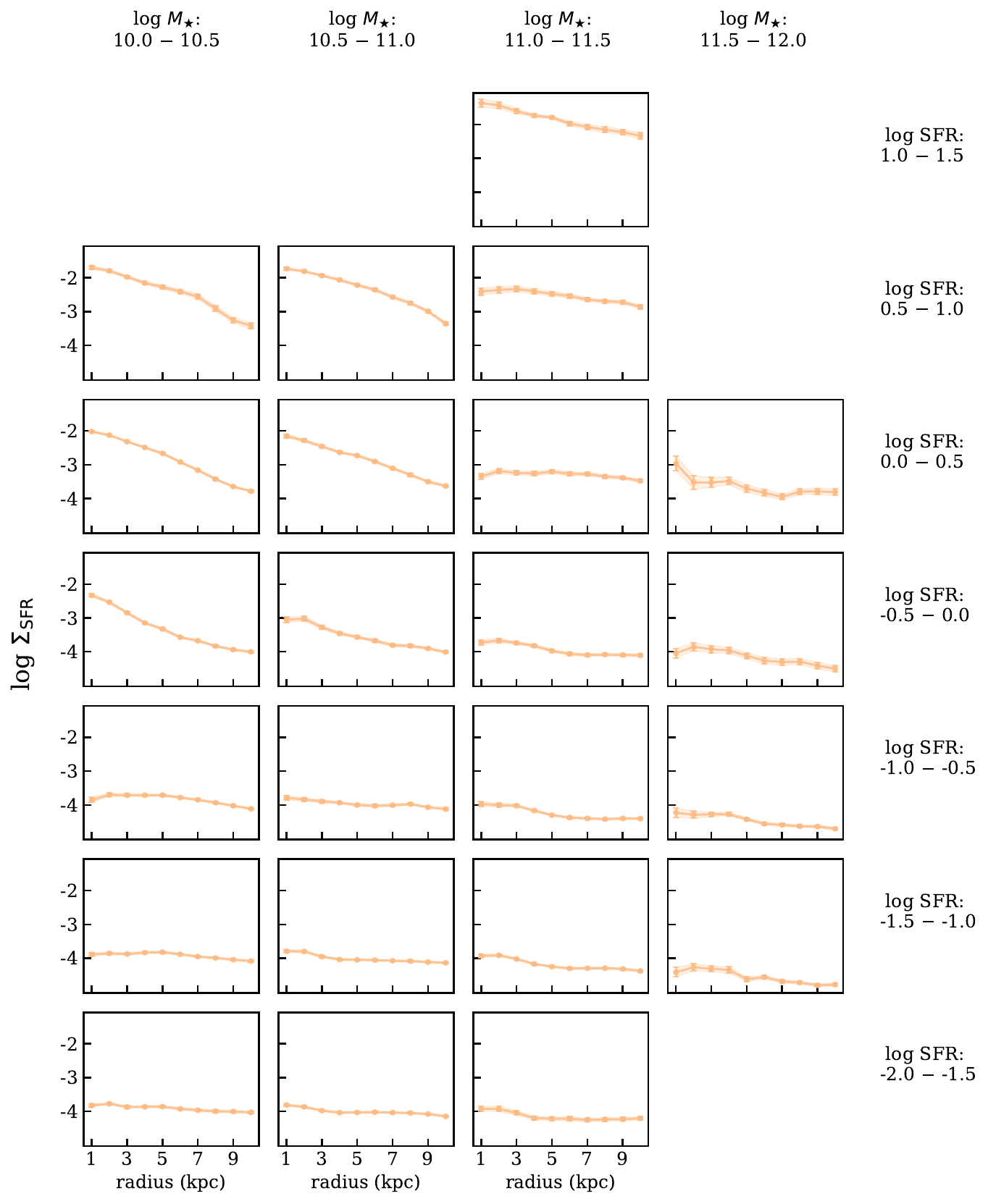}
\caption{Mean \sigmasfr\ profiles of galaxies in the sample grouped according to their \mstar\ and SFR values. Error bars represent the standard
error of the mean.}
\label{all_sfrprofile}
\end{center}
\end{figure*}

Even when the control galaxies are selected, if one does not take into account the limitation of optical AGN selection bias, suppressed central SFR can be observed among some subsamples \citep[e.g.][]{Bing2019, Lammers2023, Gatto2025}.
Also, the effects of high global SFR and high central SFR in predicting the high level of AGN activity are hard to disentangle when control galaxies are selected only based on \mstar.
In Figure~\ref{all_sfrprofile}, we plot the mean \sigmasfr\ profiles for all the galaxies in the sample, grouped according to \mstar\ and SFR values.
We can see that for a given \mstar, when the global SFR is high, the profiles are more centrally concentrated. This effect is more prominent among log \mstar\ < 11 galaxies, where the \sigmasfr\ profiles become flatter when the global SFR decreases, and \sigmasfr\ values in the central regions go through much larger variations compared to \sigmasfr\ in the outer regions.
This is consistent with the expectation of cold gas supply distributed according to the host-galaxy potential well.
Among the most massive galaxies, we can see signs of elevated \sigmasfr\ in the central regions that do not relate to fuelling at the global scale.
Together, these factors point to the need to select control galaxies with similar SFR value rather than simply controlling for \mstar\ when studying how the central star formation property, in particular, relates to AGN activity: if one does not control SFR \citep[e.g.][]{Bing2019, Mulcahey2022, Lammers2023, minijpas2024, Gatto2025}, there might be a higher risk of observing artificially suppressed \sigmasfr\ among AGN hosts due to the selection effects as we previously discussed. In our work, we also observe the decreasing optical AGN fraction at the high \sigmasfrone\ end, but since we control for the global SFR when comparing the \sigmasfr\ profiles (the significance of the selection effect varies as a function of both \sigmasfrone\ and total SFR), the high central \sigmasfr\ of optical AGNs compared to normal galaxies can also be observed (though the true difference might be larger, as there are still missing optical AGNs expected among control galaxies).                                                                                                                                                                                                                     
Furthermore, by controlling for the global SFR as well as \mstar\ through the nearest neighbour method, we can reliably observe how AGN activity is connected to the availability of gas supply in the center of the galaxy (or the distribution of the gas supply) when the total global gas supply is comparable -- rather than simply more gas in the center when there is more available throughout the whole galaxy.

We therefore advise caution in interpreting previous studies that yield different results in central star formation differences when comparing AGNs with different luminosities to normal galaxies.
First, as we discussed in Section~\ref{ss-commonfuel}, emission-line luminosity cannot safely probe AGN luminosity among low-to-moderate luminosity AGNs, which is also a challenge for radio luminosity measurements \citep[e.g.][]{Mulcahey2022, Gatto2025}. Estimation from the X-ray emission, as adopted in the study, is relatively safer, but only when the contamination from the XRB is small. 
This is also the limitation of the X-ray selection, as the nature of galaxies with log \lx\ $< 41$ would be hard to justify, and results established for this type of subsamples \citep[e.g.][]{minijpas2024} should be treated with caution.\footnote{When high-angular-resolution X-ray measurements are available, it is plausible to identify low-power X-ray AGNs more reliably as the offset of the X-ray source to the galactic nucleus can be constrained more precisely. However, current high-angular-resolution X-ray observations from \chandra\ have limited sky coverage.}
Second, similar luminosity cannot represent a similar AGN accretion level when probing galaxies with a large range of BH mass, and adding the normalization from \mstar\ can better approximate the specific accretion rate. 
Last but not least, we caution that different mean \sigmasfr/\sigmassfr\ profiles of AGNs with different luminosity levels (which is commonly adopted in previous studies) or even specific accretion rates do not necessarily mean that their host galaxies are different. 
Assuming a relatively similar accretion rate distribution throughout the galaxy population \citep[e.g.][]{Aird2019}, one might find AGNs with higher luminosity/accretion rate have higher \sigmasfrone\ among a limited-size sample, given that the probability of detecting such an AGN among low \sigmasfrone\ galaxies becomes very small.
It is plausible that the specific accretion rate distribution of AGNs varies with \sigmasfrone, but due to the limited sample size of the work, we cannot probe it confidently, and given the results from \citet{Aird2019} that the average specific accretion rate does not vary significantly with SFR at low redshift, which correlates with \sigmasfrone, one might not expect dramatic variation of the specific accretion rate distribution across different \sigmasfrone\ (except for extreme \sigmasfrone\ values).

\section{Conclusions} \label{s-c}
Utilizing the IFU spectroscopy from SDSS MaNGA, we studied how X-ray and optical AGN fractions vary as a function of \sigmasfrone, and investigated the spatially resolved properties of X-ray and optical AGN hosts compared to normal galaxies.
The main points from this paper can be summarized as follows:
\begin{itemize}
\item We built a sample of MaNGA galaxies that have X-ray coverage from eROSITA, \xmm, or \chandra, and identified X-ray AGNs (see Section~\ref{ss-mangaxray}) and optical AGNs (see Section~\ref{ss-oagn}) among this sample.
For all the objects in the sample, we constructed \sigmam, \sigmasfr, \dms, \dn\ maps to derive the corresponding radial profiles, and obtained the \sigmasfrone\ measurements (see Section~\ref{ss-measure-spatial}).
\item We studied the fractions of X-ray and optical AGNs as a function of \sigmasfrone\ (see Section~\ref{ss-agnf}).
We found that the fraction of log \lx/\mstar\ > 30.5 AGNs increases with \sigmasfrone, while the fraction of log \lx/\mstar\ < 30.5 AGNs decreases with \sigmasfrone\ among log \mstar\ $> 11$ galaxies where these X-ray AGNs with negligible levels of accretion can be detected (see Section~\ref{sss-xagnf}). 
The optical AGN fraction typically increases with \sigmasfrone, while there is a drop at the high \sigmasfrone\ end that is more prominent when the probed \mstar\ range is smaller (see Section~\ref{sss-oagnf}).
\item We further studied the spatially resolved profiles of X-ray/optical AGN hosts and compared with galaxies matched with similar global properties: log \mstar, log SFR, and $z$ (see Section~\ref{ss-agnprofile}). 
We found that both X-ray and optical AGNs tend to have higher central \sigmasfr, though log \lx/\mstar\ < 30.5 AGNs do not show significant differences in central \sigmasfr\ compared with normal galaxies when the method for measuring \sigmasfr\ is not sensitive to very recent star formation (see Section~\ref{sss-xagnprofile}).
\item We argue that the observed discrepancy in the results among X-ray AGNs and optical AGNs can be attributed to selection effects.
In addition to the well-known limitation of optical selection in picking up weak AGNs among quiescent galaxies, as identifying weak AGNs with BPT diagrams becomes harder when the star formation activity is stronger, one might observe a drop of optically-selected AGN fraction at the high \sigmasfrone\ end, which might be misinterpreted as negative AGN feedback (see Section~\ref{ss-optsb}).
\item The elevated level of central \sigmasfr\ among AGN hosts we observed in general is consistent with the picture of coeval growth of BHs and central parts of galaxies as a result from a common fuel supply; the difference in \sigmasfr\ is more prominent in the inner regions, which could explain why BH growth is more closely related with host-galaxy properties in the central $\sim$ kpc regions (see Section~\ref{ss-commonfuel}). While most of the massive galaxies in the local Universe have suppressed sSFR in the center (which can also be misinterpreted as evidence for negative AGN feedback), lower levels of central \sigmasfr\ of AGN hosts compared to normal galaxies on average should not be observed if the AGN and control galaxy samples are carefully selected (see Section~\ref{ss-compare}).
\item We also note that the picture of coeval growth from a common fuel supply does not contradict AGN feedback in action. For individual objects, it is possible to observe the effect of both positive and negative AGN feedback. We can also observe slight decreases (or flattening) of the mean \sigmasfr\ profiles toward the galaxy center on small scales ($\lesssim$ several kpc) among AGN hosts in our sample, consistent with the expectation of outflow-related AGN feedback. Among the most massive galaxies, this decreasing trend becomes more noticeable (in both scale and amplitude), consistent with the expectation of jet-related AGN feedback (see Section~\ref{ss-commonfuel}).

\end{itemize}

In the future, the accumulation of JWST IFU observations will help to extend the analyses to higher redshifts, 
where a population of more powerful AGNs can be probed among galaxies with more active star formation,
and the dynamic balance between fuelling and feedback in the common gas reservoir will have larger variations, so that both effects can be observed more clearly.
Combining jet and outflow properties of AGNs into this type of analysis will also help to test the scenario proposed in this study.

\begin{acknowledgements}
We thank the anonymous referee for constructive feedback.
This work is based on data from eROSITA, the soft X-ray
instrument aboard SRG, a joint Russian-German science mission supported by
the Russian Space Agency (Roskosmos), in the interests of the Russian Academy
of Sciences represented by its Space Research Institute (IKI), and the Deutsches
Zentrum für Luftund Raumfahrt (DLR). The SRG spacecraft was built by Lav
ochkin Association (NPOL) and its subcontractors, and is operated by NPOL
with support from the Max Planck Institute for Extraterrestrial Physics (MPE).
The development and construction of the eROSITA X-ray instrument was led
by MPE, with contributions from the Dr. Karl Remeis Observatory Bamberg
\& ECAP (FAU Erlangen-Nuernberg), the University of Hamburg Observatory,
the Leibniz Institute for Astrophysics Potsdam (AIP), and the Institute for Astronomy
and Astrophysics of the University of Tübingen, with the support of
DLR and the Max Planck Society. The Argelander Institute for Astronomy of
the University of Bonn and the Ludwig Maximilians Universität Munich also
participated in the science preparation for eROSITA. The eROSITA data shown
here were processed using the eSASS/NRTA software system developed by the
German eROSITA consortium. 
\end{acknowledgements}

\bibliographystyle{aa} 
\bibliography{sfrc_bib}

\end{document}